\documentclass[a4paper,fleqn]{cas-sc}

\usepackage{amssymb}
\usepackage{amsmath}
\usepackage{makecell}

\usepackage{lineno}

\usepackage{xcolor}
\usepackage{soul}
\usepackage{cancel}
\makeatletter
\newcommand{\linethrough}{\mathpalette\@thickbar}
\newcommand{\@thickbar}[2]{{#1\mkern0mu\vbox{
    \sbox\z@{$#1#2\mkern-1.5mu$}%
    \dimen@=\dimexpr\ht\tw@-\ht\z@+2\p@\relax 
    \hrule\@height0.5\p@ 
    \vskip\dimen@
    \box\z@}}
}
\makeatother

\usepackage{hyperref}
\hypersetup{
    colorlinks=true,     
    urlcolor=blue,
    }
\usepackage{subcaption}
\usepackage[export]{adjustbox}

\usepackage[capitalise]{cleveref} 
\usepackage[numbers, square, sort&compress]{natbib}

\usepackage{url}
\creflabelformat{equation}{#2\textup{#1}#3}

\usepackage{booktabs}
\usepackage{multirow}
\usepackage{threeparttable}
\usepackage{placeins}

\usepackage{graphicx}
\usepackage{epstopdf} 

\AppendGraphicsExtensions{.tiff}

\AppendGraphicsExtensions{.tif}

\usepackage{nomencl}
\usepackage{mdframed}
\usepackage{multicol}

\usepackage{etoolbox}
\renewcommand\nomgroup[1]{%
  \item[\bfseries
  \ifstrequal{#1}{A}{\textit{Abbreviations}}{%
  \ifstrequal{#1}{S}{\textit{Symbols}}{%
  }}%
]}

\makenomenclature
\renewcommand*\nompreamble{\begin{multicols}{2}}
\renewcommand*\nompostamble{\end{multicols}}

\def\tsc#1{\csdef{#1}{\textsc{\lowercase{#1}}\xspace}}
\tsc{WGM}
\tsc{QE}

\begin{document}
\let\WriteBookmarks\relax
\def\floatpagepagefraction{1}
\def\textpagefraction{.001}

\shorttitle{An experimental evaluation of the interplay between geometry and scale on cross-flow turbine performance}  

\shortauthors{Hunt \textit{et al.}}  

\title [mode = title]{An experimental evaluation of the interplay between geometry and scale on cross-flow turbine performance} 

\affiliation[inst1]{organization={Department of Mechanical Engineering, University of Washington},
            addressline={Mailbox 352600}, 
            city={Seattle},
            postcode={98195}, 
            state={WA},
            country={United States}}

\affiliation[inst2]{organization={XFlow Energy},
            addressline={722 S Monroe St}, 
            city={Seattle},
            postcode={98108}, 
            state={WA},
            country={United States}}

\affiliation[inst3]{organization = {California Institute of Technology},
            addressline={1200 E California Blvd},
            city={Pasadena},
            postcoce={91125},
            state={CA},
            country={United States}}

\affiliation[inst4]{organization={University of New Hampshire},
            addressline={Chase Ocean Engineering Laboratory}, 
            city={Durham},
            postcode={03824}, 
            state={NH},
            country={United States}}

\author[inst1]{Aidan Hunt}
\ead{ahunt94@uw.edu}
\cormark[1]
\cortext[1]{Corresponding author.}
\credit{Software, Validation, Formal analysis, Investigation, Writing - Original Draft and Review \& Editing, Visualization}

\author[inst1,inst2]{Benjamin Strom}
\credit{Conceptualization, Methodology, Investigation, Writing - Original Draft and Review \& Editing}

\author[inst1]{Gregory Talpey}
\credit{Validation, Formal analysis, Data Curation, Investigation, Writing - Original Draft and Review  \&  Editing, Visualization}

\author[inst1]{Hannah Ross}
\credit{Investigation, Writing - Review  \&  Editing}

\author[inst1,inst3]{Isabel Scherl}
\credit{Investigation, Writing - Review  \&  Editing}

\author[inst1]{Steven Brunton}
\credit{Supervision, Writing - Review  \&  Editing}

\author[inst4]{Martin Wosnik}
\credit{Resources, Writing - Review \& Editing}

\author[inst1]{Brian Polagye}
\credit{Conceptualization, Methodology, Resources, Writing - Original Draft and Review  \&  Editing, Supervision, and Funding acquisition}

\begin{abstract}
    Cross-flow turbines harness kinetic energy in wind or moving water.
    Due to their unsteady fluid dynamics, it can be difficult to predict the interplay between aspects of rotor geometry and turbine performance.
    This study considers the effects of three geometric parameters: the number of blades, the preset pitch angle, and the chord-to-radius ratio.
    The relevant fluid dynamics of cross-flow turbines are reviewed, as are prior experimental studies that have investigated these parameters in a more limited manner.
    Here, 223 unique experiments are conducted across an order of magnitude of diameter-based Reynolds numbers ($\approx 8\!\times\!10^4 - 8\!\times\!10^5$) in which the performance implications of these three geometric parameters are evaluated.
    In agreement with prior work, maximum performance is generally observed to increase with Reynolds number and decrease with blade count.
    The broader experimental space clarifies parametric interdependencies; for example, the optimal preset pitch angle is increasingly negative as the chord-to-radius ratio increases.
    As these experiments vary both the chord-to-radius ratio and blade count, the performance of different rotor geometries with the same solidity (the ratio of total blade chord to rotor circumference) can also be evaluated. Results demonstrate that while solidity can be a poor predictor of maximum performance, across all scales and tested geometries it is an excellent predictor of the tip-speed ratio corresponding to maximum performance.
    Overall, these results present a uniquely holistic view of relevant geometric considerations for cross-flow turbine rotor design and provide a rich dataset for validation of numerical simulations and reduced-order models.
\end{abstract}

\begin{graphicalabstract}
    \centering
    \includegraphics[width=\textwidth]{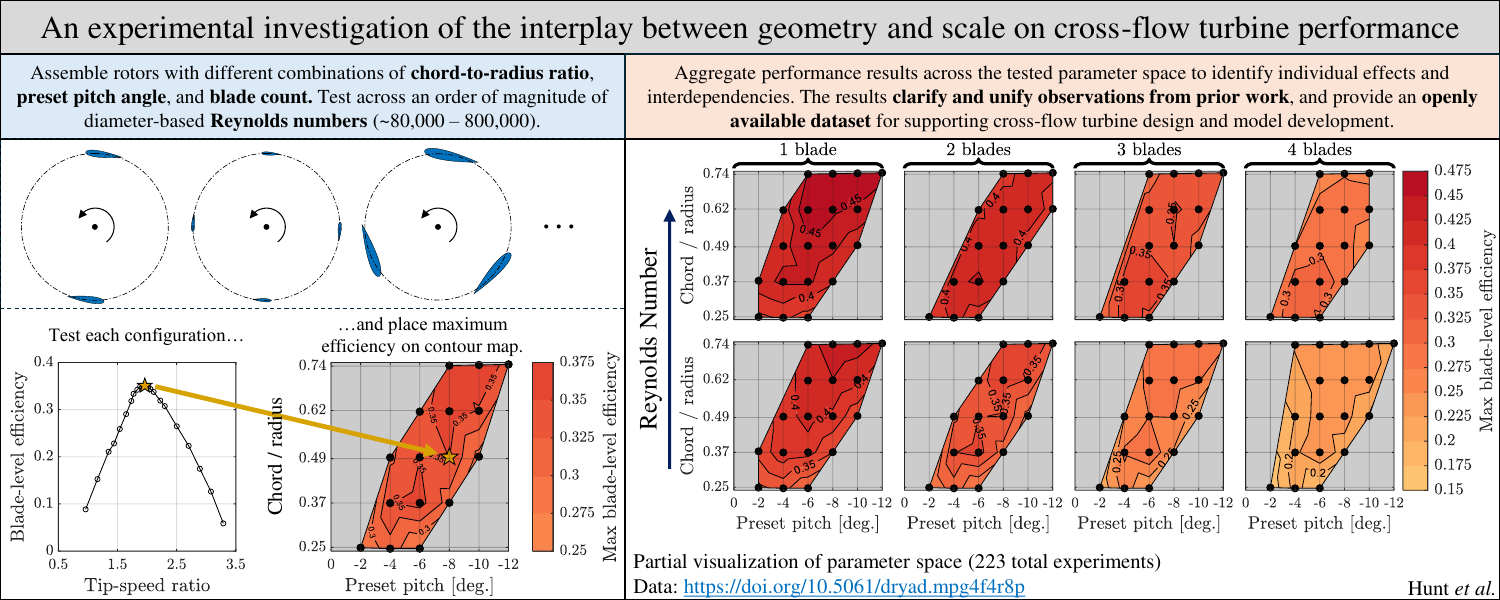}
\end{graphicalabstract}

\begin{highlights}
    \item Experimental parameter space unifies narrower prior work on cross-flow turbines
    \item Optimal chord-to-radius ratio decreases as diameter-based Reynolds number increases
    \item Optimal preset pitch angle becomes more negative as chord-to-radius ratio increases
    \item Turbines with more blades have lower efficiency, but more uniform power and forces
    \item Turbines with the same solidity do not always have the same maximum efficiency
\end{highlights}



\nomenclature[S]{$a$}{Axial induction factor}
\nomenclature[S]{$a_x$}{Streamwise induction factor}
\nomenclature[S]{$a_y$}{Cross-stream induction factor}
\nomenclature[S]{$a'$}{Angular induction factor}
\nomenclature[S]{$U_{x}$}{Streamwise velocity [m/s]}
\nomenclature[S]{$U_{y}$}{Cross-stream velocity [m/s]}

\nomenclature[S]{$\alpha$}{Angle of attack [deg.]}
\nomenclature[S]{$\alpha_n$}{Nominal angle of attack [deg.]}
\nomenclature[S]{$U_{\infty}$}{Freestream velocity [m/s]}
\nomenclature[S]{$U_{\mathrm{rel}}$}{Relative velocity on blade [m/s]}
\nomenclature[S]{$U_{\mathrm{rel},n}$}{Nominal relative velocity on blade [m/s]}

\nomenclature[S]{$A$}{Turbine projected area [m$^2$]}
\nomenclature[S]{$\alpha_p$}{Preset pitch angle [deg.]}
\nomenclature[S]{$c$}{Blade chord length [m]}
\nomenclature[S]{$c/R$}{Chord-to-radius ratio}
\nomenclature[S]{$D$}{Turbine diameter (at blade $c/4$) [m]}
\nomenclature[S]{$N$}{Number of blades}
\nomenclature[S]{$H$}{Blade span [m]}
\nomenclature[S]{$R$}{Turbine radius (at blade $c/4$) [m]}
\nomenclature[S]{$R'$}{Outermost radius swept by blades [m]}
\nomenclature[S]{$\sigma$}{Solidity}

\nomenclature[S]{$\beta$}{Blockage ratio [\%]}
\nomenclature[S]{$Fr_h$}{Depth-based Froude number}
\nomenclature[S]{$Fr_s$}{Submergence-based Froude number}
\nomenclature[S]{$g$}{Gravitational acceleration [m/s$^2$]}
\nomenclature[S]{$h$}{Dynamic water depth [m]}
\nomenclature[S]{$\nu$}{Kinematic viscosity [m$^2$/s]}
\nomenclature[S]{$\rho$}{Density [kg/m$^3$]}
\nomenclature[S]{$Re_c$}{Chord-based Reynolds number}
\nomenclature[S]{$Re_{c,n}$}{Nominal chord-based Reynolds number}
\nomenclature[S]{$Re_D$}{Diameter-based Reynolds number}
\nomenclature[S]{$s$}{Submergence depth [m]}
\nomenclature[S]{$w$}{Flume width [m]}

\nomenclature[S]{$C_L$}{Lateral force coefficient}
\nomenclature[S]{$C_P$}{Performance coefficient}
\nomenclature[S]{$C_{P,\mathrm{blades}}$}{$C_P$ of blades only}
\nomenclature[S]{$C_{P,\mathrm{supports}}$}{$C_P$ of support structures only}
\nomenclature[S]{$C_T$}{Thrust coefficient}
\nomenclature[S]{$\lambda$}{Tip-speed ratio}
\nomenclature[S]{$\lambda_{\mathrm{opt}}$}{Optimal tip-speed ratio}
\nomenclature[S]{$L$}{Lateral force [N]}
\nomenclature[S]{$\omega$}{Angular velocity [rad/s]}
\nomenclature[S]{$Q$}{Hydrodynamic torque [N-m]}
\nomenclature[S]{$T$}{Thrust force [N]}
\nomenclature[S]{$\theta$}{Angular position [deg.]}

\nomenclature[A]{UW}{University of Washington}
\nomenclature[A]{UNH}{University of New Hampshire}
\nomenclature[A]{TI}{Turbulence intensity [\%]}

\nomenclature[S]{$X$}{Component measurement of a performance quantity}
\nomenclature[S]{$X_i$}{The average value of $X$ over the $i$\textsuperscript{th} rotational cycle}
\nomenclature[S]{$\langle X \rangle$}{The time-average value of $X$ over all cycles}
\nomenclature[S]{$b_X$}{Systematic standard uncertainty of $X$}
\nomenclature[S]{$s_X$}{Random standard uncertainty of $X$}
\nomenclature[S]{$u_X$}{Combined standard uncertainty of $X$}
\nomenclature[S]{$\kappa_j$}{Sensitivity of a performance metric to the $j$\textsuperscript{th} component measurement.}

\maketitle

\begin{table*}[!t]
  \begin{mdframed}[linecolor=black]
    \printnomenclature[1.2cm]
  \end{mdframed}
\end{table*}

\section{Introduction}
\label{sec:introduction}

\begin{figure*}[t]
    \centering
    \includegraphics[width=\textwidth]{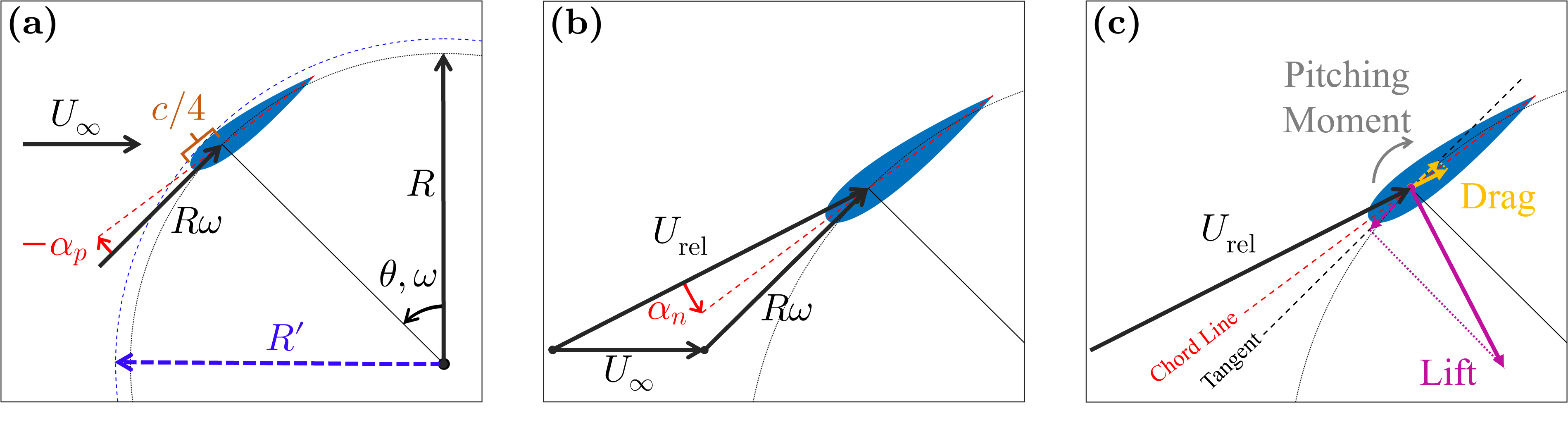}
    {\phantomsubcaption\label{fig:vec_dia_a}
    \phantomsubcaption\label{fig:vec_dia_b}
    \phantomsubcaption\label{fig:vec_dia_c}
    }
    \caption{\subref{fig:vec_dia_a} The freestream velocity, blade tangential velocity, and preset pitch angle yield \subref{fig:vec_dia_b} a relative velocity on the blade at an angle of attack $\alpha_n$, which results in \subref{fig:vec_dia_c} lift and drag forces on the foil. The projection of these forces onto the line tangent to rotation gives rise to torque on the turbine, with the projection of lift producing torque in the direction of rotation and the projection of drag opposing rotation. The representation in this figure neglects induction effects, which alter the $U_{\infty}$ and $R\omega$ vectors. Additionally, if the lift and drag forces do not act at the quarter chord of the blade (which can occur even for symmetric foils due to dynamic stall or flow curvature effects), a pitching moment about the quarter chord also contributes to the net torque on the turbine.}
    \label{fig:vec_dia}
\end{figure*}

Cross-flow turbines can be used to harness the kinetic energy in moving fluids and convert it to renewable, mechanical energy \citep{Paraschivoiu_wind_2002}. Lift-based cross-flow turbines consist of one or more blades with a foil cross-section that rotate about an axis perpendicular to the direction of fluid motion (\Cref{fig:vec_dia}). In wind energy, cross-flow turbines are typically oriented vertically and often referred to as ``vertical-axis'' turbines \citep{sutherland_retrospective_2012, mollerstrom_historical_2019}, while in water, they are often deployed in both horizontal and vertical orientations. 

If blade pitch---the angle between the blade chord line and line tangent to the rotation plane at the quarter-chord ($\alpha_p$ as shown in \Cref{fig:vec_dia_a})---is fixed, rotation rate is the single degree of freedom for the turbine during operation. Despite this mechanical simplicity, during a single rotational cycle the direction and magnitude of the inflow relative to the blade chord changes continuously, giving rise to periodically-varying dynamics. As a result, the turbine blades may experience a number of fluid phenomena that are absent for axial-flow (``horizontal-axis'') turbines, including flow curvature \citep{migliore_flow_1980}, dynamic stall \citep{laneville_dynamic_1986,simao_ferreira_visualization_2009,buchner_dynamic_2018,le_fouest_dynamic_2022, dave_analysis_2023}, and interaction between the downstream blade(s) and the wake generated by the upstream blade(s).

Due to the variation in relative velocity, lift and drag forces acting on each rotor blade vary with azimuthal position, $\theta$. The tangential component of these forces produces torque for power generation (\Cref{fig:vec_dia_c}), while their projection into Cartesian coordinates produces forces in the streamwise and cross-stream directions. Rotor torque and forces depend on multiple factors, five of which are considered in this study: two related to the flow over the turbine blades, and three related to rotor geometry.

First, for a constant rotation rate, the turbine's operating condition is defined by its tip-speed ratio,
\begin{equation}
    \lambda = \frac{R \omega}{U_\infty},
\end{equation}
\noindent where $R$ is the turbine radius, $\omega$ is the rotation rate, and $U_\infty$ is the undisturbed inflow velocity. Since blade kinematics are conventionally evaluated at the quarter-chord point, $R$ is defined here as the distance from the axis of rotation to the quarter-chord position along the chord line. However, the outermost radius swept by the turbine ($R'$) is larger and varies with both blade thickness and preset pitch angle (\Cref{fig:vec_dia_a}).

Second, the lift and drag forces on each blade depend on the local Reynolds number, $Re_c$ \citep{burton_wind_2011}
\begin{equation}
    Re_c = \frac{U_\mathrm{rel} c}{\nu},
    \label{eq:ReC}
\end{equation}
\noindent where $U_\mathrm{rel}$ is the flow velocity relative to the moving blade (\Cref{fig:vec_dia_b}), $c$ is the blade chord length, and $\nu$ is the kinematic viscosity of the surrounding fluid. Since $U_\mathrm{rel}$ depends on the velocity induced by the turbine forces and torque, it is not generally a known quantity in experiments. However, because the induced velocities depend on the turbine operating state, specifying $\lambda$, $U_\infty$, and $\nu$ for a fixed turbine geometry implicitly specifies $Re_c$ regardless of the velocity scale used in the calculation (e.g., $R \omega$, $U_\infty$). Consequently, the Reynolds number is adequately expressed by any readily available length and velocity scale (e.g., \citep{bachant_effects_2016, miller_verticalaxis_2018}). For convenience, a diameter-based Reynolds number is often employed, which is defined using the freestream velocity, $U_{\infty}$, and turbine diameter, $D = 2R$, as 
\begin{equation}
    Re_D = \frac{U_{\infty} D}{\nu}.
    \label{eq:ReD}
\end{equation}
\noindent As $Re_c$ is implicitly specified by $Re_D$ and $\lambda$, but indeterminate, it is noted that this can complicate the physical interpretation of performance differences between turbines when geometry and kinematics are simultaneously varied. 

Third, turbine torque and forces depend on the number of blades ($N$) \citep{mcadam_experimental_2013, mcadam_experimental_2013a, li_effect_2015, araya_transition_2017, miller_solidity_2021}. Fourth, torque and forces are affected by the ``preset pitch angle'', $\alpha_p$ (\Cref{fig:vec_dia_a}), which introduces an asymmetry in the direction of $U_\mathrm{rel}$ over a rotational cycle, even in the absence of any induced velocities \citep{klimas_effects_1981, takamatsu_effects_1985, fiedler_blade_2009, zhao_hydrodynamic_2013, strom_consequences_2015}. Fifth, the ``chord-to-radius ratio'', $c/R$, affects the degree of flow curvature over the blade profile. When $c$ is an appreciable fraction of $R$, the blade tangential speed varies from leading edge to trailing edge, creating ``virtual camber'' (even for symmetrical foil cross-sections) as well as ``virtual incidence" \citep{migliore_flow_1980, takamatsu_study_1985}. In addition to the geometric parameters explored here, foil profile, \citep{migliore_comparison_1983, bianchini_design_2015, ouro_effect_2018, du_experimental_2019}, helix or cant angle (i.e., blades that are twisted or tilted relative to the axis of rotation) \citep{shiono_output_2002, armstrong_flow_2012, saini_review_2019}, Coriolis effects \citep{tsai_coriolis_2016}, and surface roughness \citep{howell_wind_2010,priegue_influence_2017} can also affect performance. 

Parametric interdependence between these factors presents several challenges to exploring the fluid dynamics of cross-flow turbines, as well as to practitioners seeking an optimal turbine design. For example, one can combine $N$ and $c/R$ into a single term that describes the ``solidity'', $\sigma$, of a turbine rotor as
\begin{equation}
    \sigma = \frac{N}{2 \pi} \frac{c}{R}.
    \label{eq:solidity}
\end{equation}
However, because inverse variation of $c/R$ and $N$ can hold $\sigma$ constant, one value of $\sigma$ can correspond to different rotor geometries with different dynamics. Similarly, when changing $c$, both the local Reynolds number and flow curvature are simultaneously affected. As a result, careful study design in both experiment and simulation is required to isolate the effects of a parameter of interest, as well as to identify the effects of interactions between parameters.

As discussed in \Cref{sec:priorWork}, the effects of the Reynolds number, $c/R$, $N$, and $\alpha_p$ on cross-flow turbine performance have been explored in prior work through both experiments and simulations.
However, experimental studies are necessary to validate simulation results, and few experimental studies have parametrically evaluated enough combinations of the Reynolds number, $c/R$, $N$, and $\alpha_p$ to identify both individual and combined effects.
Furthermore, differences in experimental approaches inhibit detailed cross-comparisons between previous experimental studies.
Consequently, a single experimental investigation of cross-flow turbine performance that evaluates numerous combinations of these parameters, clarifies and unifies trends observed in prior work, and provides an expansive dataset for the validation of reduced-order and numerical models is warranted.

In this work, the interplay between cross-flow turbine rotor geometry and scale is experimentally investigated through parametric variation of the tip-speed ratio, Reynolds number, $c/R$, $N$, and $\alpha_p$. In \Cref{sec:background} the salient fluid dynamics for cross-flow turbines are highlighted, and prior experimental research involving narrower investigation of the parameters of interest is reviewed. \Cref{sec:methods} presents the methodology for the experiments, \Cref{sec:results} presents the main trends observed across the parameter space and compares these to prior work, and \Cref{sec:discussion} discusses the fluid dynamics that govern these trends. The study concludes with a summary of the implications of these results for turbine designers and future research in \Cref{sec:conclusion}.

\section{Background}
\label{sec:background}
\subsection{Cross-flow Turbine Fluid Dynamics}

\begin{figure}[t]
    \centering
    \includegraphics[width=0.5\textwidth]{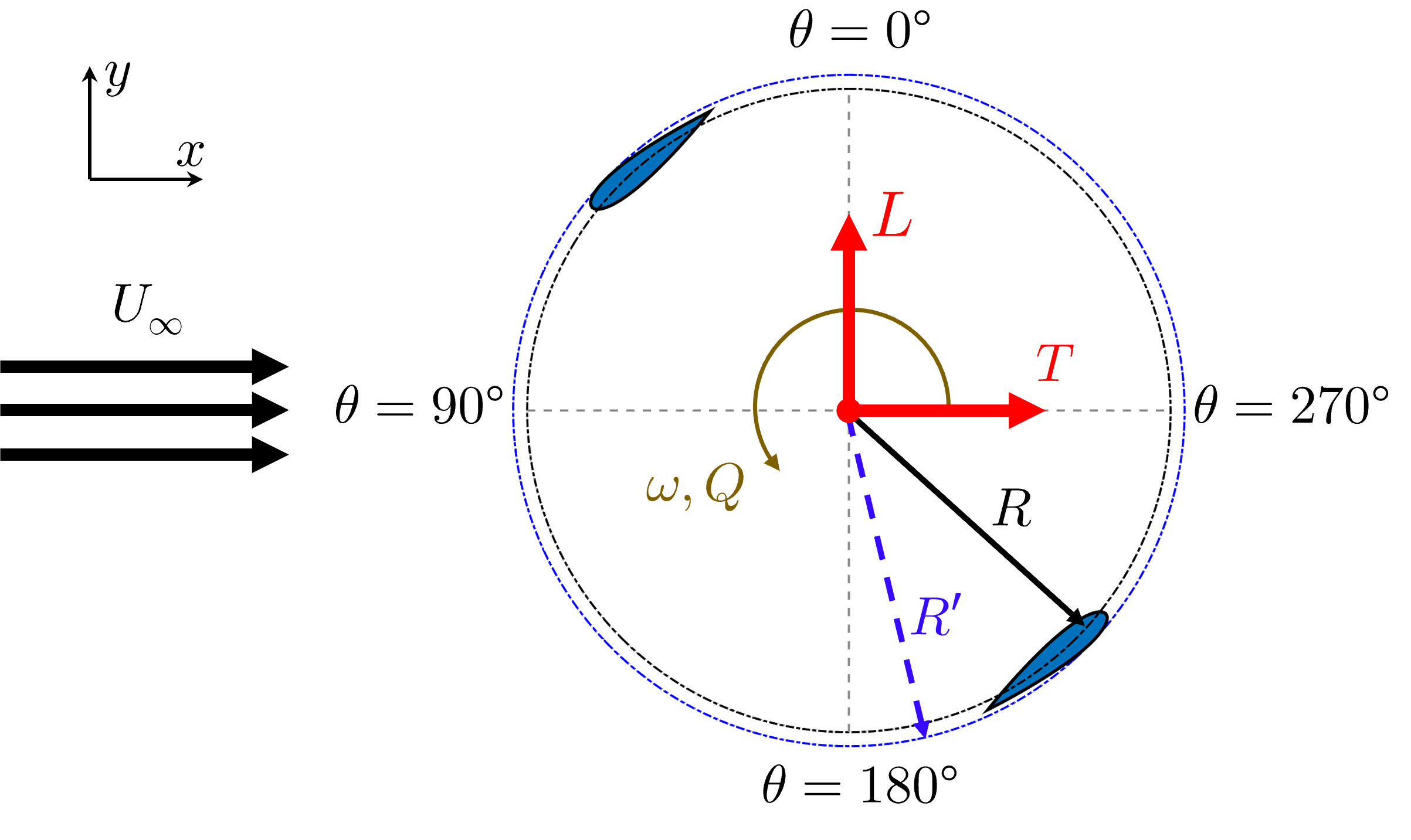}
    \caption{Overhead view of a two-bladed turbine with forces and torques annotated.}
    \label{fig:turbOverhead}
\end{figure}

The forces and torque on the turbine blades during rotation result in a net thrust force ($T$), a net lateral force ($L$), and a net fluid torque ($Q$) on the turbine as shown in \Cref{fig:turbOverhead}. As the cross-flow turbine harvests power from a fluid stream, these forces and torques must be balanced by a change in momentum in the freestream. 
For forces, to satisfy linear momentum conservation, an alteration to the inflow velocity as power is extracted is required. For axial-flow turbines, where the sole significant force is thrust in the streamwise direction, this is described by an ``axial induction'' factor, $a$, which represents the fractional decrease of freestream velocity at the rotor plane and is optimally equal to 1/3 \citep{burton_wind_2011}. 
Since cross-flow turbines experience net force in both the streamwise and cross-stream directions, induction from these forces could be described by a pair of ``linear induction'' factors, $a_x$ and $a_y$, respectively, such that $U_x = U_\infty(1-a_x)$ and $U_y = -a_y U_\infty$. Since blade forces vary with azimuthal position, $\theta$, these factors would as well (i.e., $a_x = a_x(\theta)$).
Similarly, for both axial-flow and cross-flow turbines, when torque is generated by the turbine, angular momentum conservation gives rise to an ``angular induction'' factor, $a'$, which optimally approaches zero at sufficiently high rotation rates \citep{burton_wind_2011}. 
Together, these factors alter both the magnitude and direction of $U_{\mathrm{rel}}$.

\begin{figure}[t]
    \centering
    \includegraphics[width=0.4\textwidth]{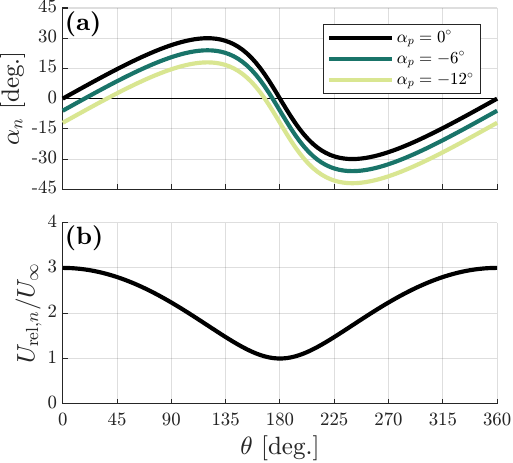}
    
    {\phantomsubcaption\label{fig:nomAlpha}
     \phantomsubcaption\label{fig:nomURel}}
    \caption{\subref{fig:nomAlpha} Nominal angle of attack, $\alpha_n$, as a function of azimuthal position during a rotational cycle for $\lambda = 2$ and several $\alpha_p$. \subref{fig:nomURel} Nominal relative velocity, normalized by $U_\infty$ for the same azimuthal positions.}
    \label{fig:alpha_Urel}
\end{figure}

Using these definitions, the variation in angle of attack, $\alpha$, throughout the blade's revolution is completely described as
\begin{equation}
    \alpha = \arctan \left( \frac{U_x\sin\theta + U_y\cos\theta}{R\omega(1+a')+U_x\cos\theta+U_y\sin\theta} \right) + \alpha_p \ \ ,
    \label{eq:alpha}   
\end{equation}

\noindent where the preset pitch angle, $\alpha_p$, is defined as in \Cref{fig:vec_dia_a} with $\alpha_p < 0^{\circ}$ corresponding to a ``toe out'' (i.e., leading edge out) rotation that reduces the local angle of attack on the upstream sweep ($0^{\circ} \leq \theta < 180^{\circ}$). Similarly, the relative velocity, $U_\mathrm{rel}$, is defined as the vector sum of the normal and tangential velocities as
\begin{equation}
    \begin{aligned}
        U_{\mathrm{rel}}= \sqrt{\left( U_x\sin\theta+U_y\cos\theta \right)^2 + \left( R\omega(1+a') + U_x\cos\theta + U_y\sin\theta \right)^2}.
    \end{aligned}
    \label{eq:U_rel}
\end{equation}
\noindent Unlike axial-flow turbines, there is not an analytical theory that generally describes $a_x$, $a_y$, and $a'$. The closest analog, Double Multiple Streamtube Theory (DMST) \citep{strickland_darrieus_1975,bedon_evaluation_2014}, estimates $a_x(\theta)$, but implicitly assumes $a_y(\theta) = a'(\theta) = 0$. Consequently, in semi-quantitative discussions of cross-flow turbine fluid mechanics, it is often convenient to neglect induction entirely and present nominal versions of $\alpha$ and $U_{\mathrm{rel}}$. If $a_x(\theta) = a_y(\theta) = a'(\theta) = 0$, then \Cref{eq:alpha,eq:U_rel} reduce to
\begin{equation}
     \alpha_n = \arctan\left(\frac{\sin\theta}{\lambda+\cos\theta}\right)+\alpha_p
    \label{eq:alpha_n}   
    \end{equation}
\noindent and
\begin{equation}
    U_{\mathrm{rel},n} = U_\infty \sqrt{\lambda^2+2\lambda\cos(\theta)+1}.
    \label{eq:U_n}
\end{equation}
As shown in \Cref{fig:alpha_Urel}, this simplified model provides some insight into cross-flow turbine dynamics throughout a rotation. First, the oscillatory nature of the forces and torque on the rotor are apparent from the variations in $\alpha_n$ and $U_{\mathrm{rel},n}$. Second, the angle of attack variation is relatively large, such that $\alpha_p < 0^{\circ}$ delays stall during the upstream sweep. Third, the maximum value for $U_{\mathrm{rel},n}$ (\Cref{fig:nomURel}) is out of phase with that of $\alpha_n$ (\Cref{fig:nomAlpha}), such that the phase-maximum for lift-generating torque is unlikely to occur at the phase-maximum for the coefficient of lift. It is important to remember that this model does not describe the actual relative velocity or angle of attack, especially in the downstream sweep ($180^{\circ} \leq \theta < 360^{\circ})$ during which the blade encounters disturbed flow. As such, direct comparisons with the static stall angle are not generally appropriate.

\subsection{Prior Experimental Work}
\label{sec:priorWork}

\begin{table*}[t]
    \centering
    \caption{Prior experimental studies of cross-flow turbines that varied the parameters considered in this work.}
    \label{tab:prior_work}
        \begin{threeparttable}
            \begin{tabular}{@{}lccccc@{}}
                \toprule
                \textbf{Reference} & \textbf{Year} & \boldmath{$N$} & \boldmath{$c/R$} & \boldmath{$\alpha_p$}   \textbf{[deg.]} & \boldmath{$Re_D \ [\times   10^{-5}]$} \\ \midrule
                \citet{blackwell_selected_1977} & 1977 & 2, 3 & 0.06 -- 0.09 & Not specified & $\sim$ 3.6 -- 41 \\
                \citet{sheldahl_aerodynamic_1980} & 1980 & 2, 3 & 0.06 & Not specified & $\sim$ 10 -- 69 \\
                Migliore et al.   \citep{migliore_flow_1980,migliore_effects_1980} & 1980 & 2 & 0.11 -- 0.26 & 0 & $\sim$ 2.5 -- 15 \\
                \citet{klimas_effects_1981} & 1981 & 2 & 0.06 & -7 -- 3 & $\sim$ 11 -- 46 \\
                \citet{worstell_aerodynamic_1981} & 1981 & 2, 3 & 0.07 & Not specified & $\sim$ 100\tnote{\dag} \\
                \citet{takamatsu_study_1985} & 1985 & 1, 2, 4 & 0.3 & 0 & 2.0 -- 6.0\tnote{\dag} \\
                \citet{takamatsu_effects_1985} & 1985 & 1 & 0.3 & -10 -- 10 & 3.0\tnote{\dag} \\
                \citet{takamatsu_experimental_1991} & 1991 & 1 & 0.20 -- 0.30 & -5 -- 0 & 3.7\tnote{\dag} \\
                \citet{ashwill_measured_1992} & 1992 & 2 & 0.07 & Not specified & $\sim$ 423\tnote{\dag} \\
                \citet{shiono_experimental_2000} & 2000 & 1 -- 3 & 0.23 -- 1.13 & 0 & 1.8 -- 4.2\tnote{\dag} \\
                \citet{fiedler_blade_2009} & 2009 & 3 & 0.32 & -8 -- 8 & 7.8 -- 17\tnote{\dag} \\
                \citet{howell_wind_2010} & 2010 & 2, 3 & 0.33 & 0 & 1.2 -- 2.0\tnote{\dag} \\
                \citet{el-samanoudy_effect_2010} & 2010 & 2 -- 4 & 0.2 -- 0.75 & -60 -- 10 & 2 -- 4\tnote{\dag} \\
                \citet{tanino_influence_2011} & 2011 & 3, 5 & 0.23 -- 0.47 & -10 -- 20\tnote{$\ddagger$} & 1.3 -- 2.4 \\
                \citet{armstrong_flow_2012} & 2012 & 3 & 0.31 & -12 -- 12 & 33 \\
                \citet{mcadam_experimental_2013,mcadam_experimental_2013a} & 2013 & 3 -- 6 & 0.26 & -4 -- 0 & 1.5 -- 2.7\tnote{\dag} \\
                \citet{zhao_hydrodynamic_2013} & 2013 & 3 & 0.24 & -3 -- 5\tnote{$\ddagger$} & 9 -- 13 \\
                \citet{li_effect_2015} & 2015 & 2 -- 5 & 0.27 & -10 -- 4 & 10\tnote{\dag} \\
                \citet{bachant_effects_2016} & 2016 & 3 & 0.28 & 0 & 3.0 -- 13 \\
                \citet{eboibi_experimental_2016} & 2016 & 3 & 0.09 -- 0.11 & 0 & 2.6 -- 3.5\tnote{\dag} \\
                \citet{li_effect_2016} & 2016 & 2 -- 5 & 0.27 & Optimal from \citep{li_effect_2015} & 10\tnote{\dag} \\
                \citet{bachant_experimental_2016} & 2016 & 3 & 0.10 & 0 & 4.3 -- 13 \\
                \citet{priegue_influence_2017} & 2017 & 2 -- 4 & 0.40 -- 0.60 & -3 & 2.9 -- 4.6\tnote{\dag} \\
                \citet{araya_transition_2017} & 2017 & 2, 3, 5 & 0.67 & 0 & 0.80 \\
                \citet{miller_verticalaxis_2018} & 2018 & 5 & 0.45 & 0 & 5 -- 50 \\
                \citet{lam_assessment_2018} & 2018 & 2, 3, 5 & 0.30 & Not specified & 2.4\tnote{\dag} \\
                \citet{somoano_dead_2018} & 2018 & 3 & 0.32 & -16 -- 8 & 3 - 5 \\
                \citet{du_experimental_2019} & 2019 & 3 & 0.22 -- 0.33 & -4 -- 0 & 2.8 -- 4.2 \\
                \citet{miller_solidity_2021} & 2021 & 2 -- 5 & 0.45 & 0 & 7 -- 50 \\
                \citet{ross_effects_2022} & 2022 & 2 & 0.49 & -6 & 0.61 -- 0.79 \\
                \citet{szczerba_wind_2023} & 2023 & 4 & 0.22 & 0 -- 4\tnote{$\ddagger$} & 2.7 -- 8.2\tnote{\dag} \\ \midrule
                Present Study &  & 1 -- 4 & 0.25 -- 0.74 & -12 -- 0 & 0.75 -- 8.3 \\ \bottomrule
            \end{tabular}
            \begin{tablenotes}
                \footnotesize{
                 \item[\dag] $Re_D$ is not reported by the reference, but estimated here using reported turbine dimensions and flow quantities. If not provided, $\nu$ is assumed to be $1.6\!\times\!10^{-5}$ for air (30 $^o$C) and $1.0\!\times\!10^{-6}$ for water (20 $^o$C).
                  \item[$\ddagger$] Sign convention for reported $\alpha_p$ is not defined by the study.
                }
            \end{tablenotes}
        \end{threeparttable}
\end{table*}

Prior experimental research varying one or more of $c/R$, $\alpha_p$, $N$, and the Reynolds number is summarized in \Cref{tab:prior_work}.
While it is acknowledged that there is a broad, complementary body of prior work utilizing numerical simulations to explore similar geometric parameters (e.g., \citep{bianchini_design_2015, rezaeiha_characterization_2018, rezaeiha_optimal_2018, hand_aerodynamic_2021}), given the necessity of experimental data for validating simulations, particularly for cases with strong dynamic stall, prior experimental work is focused on here.
Additionally, for ease of comparison across studies, the diameter-based Reynolds number(s) (\Cref{eq:ReD}) associated with each study are provided, as this definition of the Reynolds number is agnostic to blade geometry.
Except for five studies which focus on troposkein rotors \citep{blackwell_selected_1977, sheldahl_aerodynamic_1980, klimas_effects_1981, worstell_aerodynamic_1981, ashwill_measured_1992}, the studies in \Cref{tab:prior_work} utilize turbines with straight blades, although a variety of blade cross-sectional profiles are employed.
Prior work has generally varied one or more of $c/R$, $N$, $\alpha_p$, and $Re_D$, while holding the others constant.
All four parameters of interest have been shown to significantly affect cross-flow turbine performance.
In general, turbine efficiency is found to increase with Reynolds number until a threshold is reached \citep{bachant_effects_2016,miller_verticalaxis_2018,miller_solidity_2021}, decrease with increasing blade count \citep{mcadam_experimental_2013, li_effect_2015, araya_transition_2017, miller_solidity_2021}, and benefit from moderate toe-out preset pitch (i.e., $-8^\circ < \alpha_p < -2^{\circ}$) \citep{klimas_effects_1981, fiedler_blade_2009, armstrong_flow_2012, mcadam_experimental_2013, zhao_hydrodynamic_2013, li_effect_2015, somoano_dead_2018}.
There are conflicting performance trends for the chord-to-radius ratio \citep{blackwell_selected_1977, migliore_flow_1980, shiono_experimental_2000, tanino_influence_2011, eboibi_experimental_2016, priegue_influence_2017, du_experimental_2019}, though this is difficult to vary independently of other parameters (e.g., $Re_c$).

While the prior investigations listed in \Cref{tab:prior_work} span a wide range of $c/R$, $\alpha_p$, $N$, and $Re_D$, differences in approach complicate in-depth cross-comparisons between these studies.
For example, several studies that explore the influences of chord-to-radius ratio or blade count attribute the observed effects to solidity (combining $c/R$ and $N$ as in \Cref{eq:solidity}) rather than to the individual parameter that was varied \citep{blackwell_selected_1977, shiono_experimental_2000, mcadam_experimental_2013, eboibi_experimental_2016, lam_assessment_2018, du_experimental_2019, miller_solidity_2021}. 
Additionally, $\alpha_p$ depends on the chord-wise location where the angle between the chord line and the tangent line is defined \citep{takamatsu_effects_1985, fiedler_blade_2009, bianchini_influence_2016, somoano_dead_2018}. For example ``neutral pitch'' ($\alpha_p = 0^{\circ}$) referenced at the mid-chord corresponds to a toe-out pitch angle ($\alpha_p < 0^{\circ}$) referenced at the quarter-chord. However, the reference location that was used is seldom reported. Consequently, although \Cref{tab:prior_work} lists the values of $\alpha_p$ reported by the authors of each study, unstated inconsistencies in the reference location for $\alpha_p$ obfuscate direct comparison.
Lastly, the studies listed in \Cref{tab:prior_work} employ a variety of blade-strut connection schemes, which can substantially affect turbine-level performance \citep{strom_impact_2018}. These factors all likely contribute to apparently conflicting conclusions about geometric trends.

Relative to prior experimental work, this study considers the most commonly tested blade counts (as well as single-bladed performance), moderate to high chord-to-radius ratios, and toe-out preset pitch angles in the general range of those that have been found to benefit performance.
The range of Reynolds numbers ($Re_D = 0.75\!\times\!10^5 - 8.3\!\times\!10^5$) spans roughly an order of magnitude and overlaps with prior experimental work, but is still an order of magnitude lower than the highest Reynolds numbers tested to date \citep{worstell_aerodynamic_1981,ashwill_measured_1992}.
Based on prior work by Bachant et al. \citep{bachant_effects_2016, bachant_experimental_2016} and \citet{miller_verticalaxis_2018, miller_solidity_2021}, the $Re_D$ tested in this study are expected to overlap with the transitional regime in which airfoil performance---and thus turbine performance---depends strongly on the Reynolds number.
The implications of these lower Reynolds numbers for the broader applicability of the results of this study are discussed in \Cref{sec:gaps}.
In contrast to most studies in \Cref{tab:prior_work}, several combinations of the three geometric parameters and Reynolds number are considered to identify interdependencies. 
While most prior experimental work characterizes turbine power production, measurements of forces are less common. Of the 31 studies in \Cref{tab:prior_work}, only \citet{mcadam_experimental_2013, mcadam_experimental_2013a}, Bachant et al. \citep{bachant_effects_2016, bachant_experimental_2016}, and \citet{li_effect_2015} characterize thrust force, and no studies characterize lateral force (although this could be derived from normal and tangential blade forces presented by Li et al. \citep{li_effect_2015,li_effect_2016}). Consequently, the experiments in this study---in which turbine power and forces are both reported---provide the most extensive characterization of cross-flow turbine performance to date.

\section{Methods}
\label{sec:methods}

To characterize cross-flow turbine performance over an order of magnitude of $Re_D$ while holding other dimensionless parameters approximately constant, two different physical turbine sizes were tested, each in a different facility. As such, this section presents the geometric parameter space and $Re_D$ explored, a description of each test facility, the metrics used to characterize performance, and the methods employed to synthesize the data.

\subsection{Experimental Turbines}
\label{methods:expTurbs}

\begin{table}[t]
    \centering
    \caption{Geometric parameters and Reynolds numbers tested for each turbine size.}
    \label{tab:turbine_dimensions}
    \begin{threeparttable}
        \begin{tabular}{@{}ccc@{}}
            \toprule
            \textbf{Parameter} & \textbf{Turbine A} & \textbf{Turbine B} \\ \midrule
            $H$ [cm] & 23.4 & 116.6 \\
            $R$ [cm]\tnote{$\ast$} & 8.21 & 41.02 \\
            $R'$ [cm]\tnote{$\ast$} & 8.23 -- 9.13 & 41.55 -- 42.88 \\
            $c$ [cm] & 2.03, 3.045, 4.06, 5.075, 6.09 & 10.10, 15.15, 20.20 \\ \midrule
            $c/R$ & 0.25, 0.37, 0.49, 0.62, 0.74 & 0.25, 0.37, 0.49 \\
            $\alpha_p$ [deg.] & 0, -2, -4, -6, -8, -10, -12 & 2, 4, 6, 8 \\
            $N$ & 1, 2, 3, 4 & 2, 4 \\ \midrule
            $Re_D \ [\times\!10^5]$ & 0.75, 1.6, 2.7 & 8.3 \\
            Test Facility & Tyler flume (UW) & Chase towing tank (UNH) \\ \bottomrule
        \end{tabular}
        \begin{tablenotes}
             \footnotesize{
             \item[$\ast$] Due to the blade mounting scheme used, $R$ varies slightly with $\alpha_p$ for each geometry, whereas $R'$ varies with both $c/R$ and $\alpha_p$. The values of $R$ and $R'$ for all tested geometries are provided in \Cref{tab:radVar}.
             }
        \end{tablenotes}
    \end{threeparttable}
\end{table}

\begin{figure*}[t]
    \centering
    {\phantomsubcaption\label{fig:uw_turb_schematic}
    \phantomsubcaption\label{fig:unh_turb_schematic}
    \phantomsubcaption\label{fig:turb_scales_comparison}
    }
    \includegraphics[width=\textwidth]{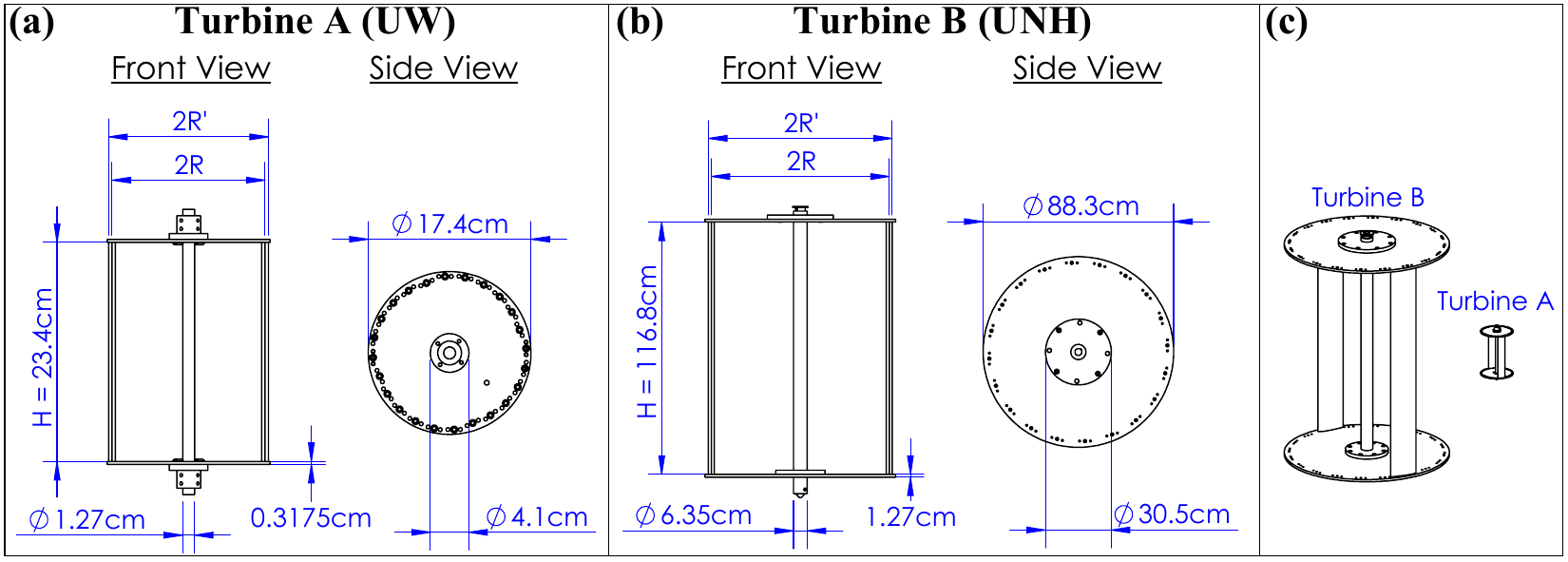}
    \caption{Dimensioned schematics of the \subref{fig:uw_turb_schematic} Turbine A and \subref{fig:unh_turb_schematic} Turbine B rotors. \subref{fig:turb_scales_comparison} Relative size comparison between the two rotors. The configurations shown correspond to $c/R = 0.49$, $N = 2$, and $\alpha_p = -6^{\circ}$.}
    \label{fig:turb_schematics}
\end{figure*}

\begin{figure*}[t]
    \centering
    \includegraphics[width=\textwidth]{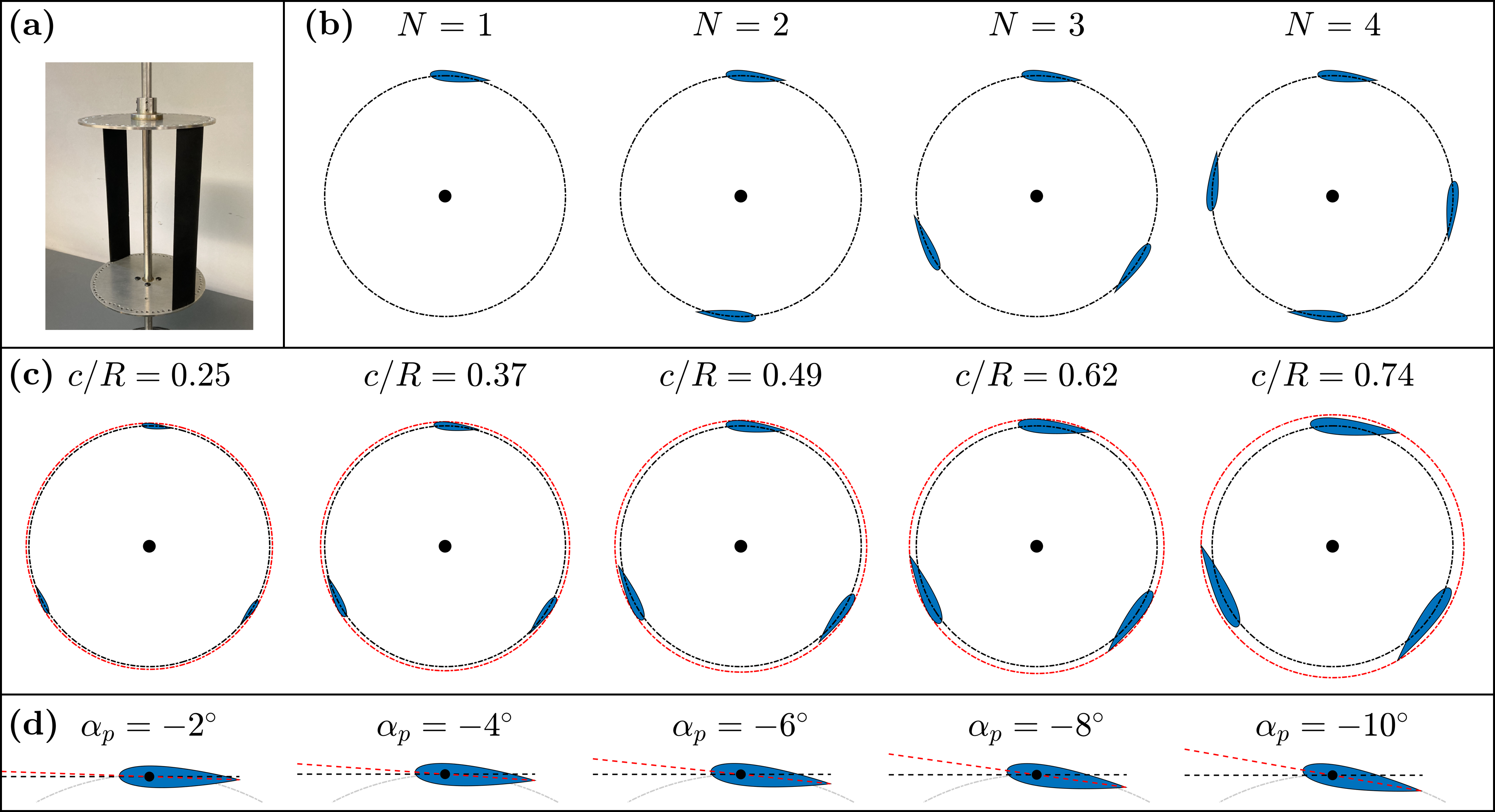}

    {\phantomsubcaption\label{fig:turbIso}
    \phantomsubcaption\label{fig:bladeVar}
    \phantomsubcaption\label{fig:chordVar}
    \phantomsubcaption\label{fig:pitchVar}
    }
    \caption{Partial visualization of experimental parameter space. \subref{fig:turbIso} Isometric view of an assembled Turbine A rotor with $N = 2$, $c/R = 0.49$, and $\alpha_p = -6^{\circ}$. \subref{fig:bladeVar} Variation in blade count, $N$, for $c/R$ = 0.49 and $\alpha_p = -6^{\circ}$. \subref{fig:chordVar} Variation in $c/R$ for $\alpha_p = -6^{\circ}$ and $N = 3$, with corresponding variation in the outermost swept radius $R'$ due to the blade mounting scheme shown in red. \subref{fig:pitchVar} Variation in $\alpha_p$ for $c/R$ = 0.49, with the chord lines indicated in red and the tangent lines indicated in black.}
    \label{fig:parameter_space}
\end{figure*}

The geometric parameter space explored in this work includes five nominal chord-to-radius ratios ($c/R$ = 0.25, 0.37, 0.49, 0.62, and 0.74), four blade counts ($N$ = 1, 2, 3, and 4), and seven preset pitch angles defined at the quarter-chord ($\alpha_p$ = $0^{\circ}, -2^{\circ}, -4^{\circ}, -6^{\circ}, -8^{\circ}, -10^{\circ}$, and $-12^{\circ}$). All other relevant turbine geometric parameters, such as the aspect ratio ($H/D$ = 1.42, where $H$ is the blade span) and blade profile (straight-blades with NACA 0018 cross-sections) were held constant. 
Two turbine sizes were tested at four $Re_D$ (\Cref{fig:turb_schematics}): the smaller Turbine A configurations (\Cref{fig:uw_turb_schematic}) were tested at $Re_D = 0.75 \times 10^5$, $1.6 \times 10^5$, and $2.7 \times 10^5$, while the larger Turbine B configurations (\Cref{fig:unh_turb_schematic}) were tested at $Re_D = 8.3 \times 10^5$.
Turbine dimensions and geometric parameters are summarized in \Cref{tab:turbine_dimensions}, and a subset of geometric configurations are visualized in \Cref{fig:parameter_space}.
For each combination of $c/R$, $N$, and $Re_D$, only the subset of $\alpha_p$ values required to identify the $\alpha_p$ that yielded maximum performance were tested.
Additionally, fewer geometric configurations were tested with Turbine B rotors than for Turbine A rotors due to the challenges of working with physically larger turbines and the realization, following experiments with the large turbines, that it would be prudent to expand the parameter space. In total, performance was characterized for 223 unique combinations of $c/R$, $\alpha_p$, $N$, and $Re_D$.

The blades for the Turbine A rotors were machined from 7075-T6 aluminum and painted with an ultra-flat black spray paint.
The Turbine B blades were constructed of carbon fiber ($c = 10.10$ cm) or fiberglass ($c = 15.15$ cm and $20.20$ cm). For both materials, blades were constructed using a wet lay-up in halves on open-faced molds and cured under vacuum in an autoclave. Blade halves were bonded together with aluminum inserts in the ends for fastening, a fiberglass dowel along the leading edge to provide additional bonding surface area, and closed-cell expanded foam core to discourage water retention.
Due to differences in manufacturing approach, the surface roughness of the Turbine A blades was likely different than that of the Turbine B blades, but was not characterized for either turbine. 

For both turbine sizes, the blades were connected to a drive shaft by circular end plates (\Cref{fig:turbIso}). While foil-shaped blade-end struts would incur lower parasitic torque losses \citep{strom_impact_2018,bachant_experimental_2016}, a unique strut assembly would be needed for each combination of $c/R$, $N$, and $\alpha_p$. To explore the parameter space while minimizing manufacturing costs, sets of end plates for each turbine size were machined with hole patterns that accommodated various numbers of equally spaced blades at various preset pitch angles. These common sets of end plates were then used for all chord lengths at each turbine scale.

This modular approach influenced the turbine design in two ways. First, for Turbine A configurations, to allow for the smallest chord lengths ($c/R = 0.25$ and $0.37$) to be compatible with the same mounting hardware used for the larger chord lengths, the chord near the blade ends was gradually increased, while the middle 92\% of the blade span retained the target chord length.
Second, the radius of the outermost circle swept by the blades ($R'$) varied for each combination of $c/R$ and $\alpha_{p}$. Specifically, because blades of all chord lengths use the same end plates, the mounting position for each preset pitch angle was selected such that the $R'$ swept by a blade with the mid-range chord length ($c = 4.06 \ \mathrm{cm}$ for Turbine A configurations; $c = 15.15$ cm for Turbine B configurations) was constant for all preset pitch angles. 
Consequently, for a blade mounted at a given $\alpha_p$, while the quarter-chord position ($R$) and chord line orientation are independent of the chord length, $R'$ changes slightly due to the variation in foil thickness and orientation. The difference in $R'$ across chord lengths for $\alpha_p = -6^{\circ}$ is shown by the red circles in \Cref{fig:chordVar}. In addition, the quarter-chord radius ($R$) varies slightly with $\alpha_p$ (maximum deviation from mean value: $\approx \! 2\%$) and to a lesser degree than $R'$ (maximum deviation from mean value: $\approx \! 6\%$). 
To account for this variation, the quarter-chord radius $R$ is used for evaluating all kinematic quantities (e.g., $\lambda$, $c/R$, $Re_D$), while $R'$ is used to calculate the turbine projected area for normalized performance metrics. The values of $R$, $R'$, and derived quantities are tabulated in \Cref{tab:radVar} for all tested geometries.

\begin{table}[t]
    \centering
    \caption{Variation in quarter-chord radius, swept radius, and derived quantities for the tested geometries due to the blade mounting scheme.}
     \begin{tabular}{@{}cccccccc@{}}
        \toprule
        \boldmath{$c$} \textbf{[cm]} & \boldmath{$\alpha_p$} \textbf{[deg.]} & \boldmath{$R$} \textbf{[cm]} & \boldmath{$R'$} \textbf{[cm]} & \boldmath{$R'/R$} & \boldmath{$c/R$} & \boldmath{$A$} \textbf{[cm$^2$]} & \boldmath{$\beta$} \textbf{[\%]} \\ \midrule
        \multirow{4}{*}{2.03} & 0 & 8.04 & 8.23 & 1.02 & 0.252 & 385.0 & 10.7 \\
         & -2 & 8.15 & 8.33 & 1.02 & 0.249 & 389.9 & 10.7 -- 10.8 \\
         & -4 & 8.24 & 8.42 & 1.02 & 0.246 & 393.9 & 10.8 -- 10.9 \\
         & -6 & 8.23 & 8.41 & 1.02 & 0.247 & 393.7 & 10.8 -- 10.9 \\ \midrule
        \multirow{4}{*}{3.045} & -2 & 8.15 & 8.42 & 1.03 & 0.374 & 394.1 & 10.9 \\
         & -4 & 8.24 & 8.51 & 1.03 & 0.370 & 398.2 & 10.9 -- 11.1 \\
         & -6 & 8.23 & 8.51 & 1.03 & 0.370 & 398.0 & 10.9 -- 11.0 \\
         & -8 & 8.21 & 8.50 & 1.04 & 0.371 & 397.8 & 11 \\ \midrule
        \multirow{5}{*}{4.06} & -2 & 8.15 & 8.60 & 1.06 & 0.498 & 402.5 & 11.1 \\
         & -4 & 8.24 & 8.60 & 1.04 & 0.493 & 402.5 & 11.0 -- 11.2 \\
         & -6 & 8.23 & 8.60 & 1.05 & 0.494 & 402.5 & 11.0 -- 11.2 \\
         & -8 & 8.21 & 8.60 & 1.05 & 0.494 & 402.5 & 11.1 -- 11.2 \\
         & -10 & 8.20 & 8.60 & 1.05 & 0.495 & 402.5 & 11.1 -- 11.2 \\ \midrule
        \multirow{6}{*}{5.075} & -2 & 8.15 & 8.87 & 1.09 & 0.623 & 415.3 & 11.5 \\
         & -4 & 8.24 & 8.83 & 1.07 & 0.616 & 413.2 & 11.4 - 11.5 \\
         & -6 & 8.23 & 8.70 & 1.06 & 0.617 & 407.0 & 11.2 -- 11.3 \\
         & -8 & 8.21 & 8.70 & 1.06 & 0.618 & 407.2 & 11.2 -- 11.3 \\
         & -10 & 8.20 & 8.71 & 1.06 & 0.619 & 407.4 & 11.2 -- 11.3 \\
         & -12 & 8.18 & 8.71 & 1.07 & 0.621 & 407.7 & 11.3 \\ \midrule
        \multirow{5}{*}{6.09} & -4 & 8.24 & 9.13 & 1.11 & 0.739 & 427.5 & 11.7 -- 11.9 \\
         & -6 & 8.23 & 8.98 & 1.09 & 0.740 & 420.3 & 11.5 -- 11.7 \\
         & -8 & 8.21 & 8.82 & 1.07 & 0.742 & 413.0 & 11.3 -- 11.5 \\
         & -10 & 8.20 & 8.82 & 1.08 & 0.743 & 412.5 & 11.3 -- 11.4 \\
         & -12 & 8.18 & 8.83 & 1.08 & 0.745 & 413.1 & 11.3 -- 11.5 \\ \midrule
        \multirow{4}{*}{10.10} & -2 & 41.52 & 42.42 & 1.02 & 0.243 & 9911.6 & 11.1 \\
         & -4 & 41.51 & 42.41 & 1.02 & 0.243 & 9910.7 & 11.1 \\
         & -6 & 41.47 & 42.40 & 1.02 & 0.244 & 9907.7 & 11.1 \\
         & -8 & 41.43 & 42.38 & 1.02 & 0.244 & 9903.5 & 11.1 \\ \midrule
        \multirow{3}{*}{15.15} & -4 & 41.51 & 42.87 & 1.03 & 0.365 & 10016.7 & 11.2 \\
         & -6 & 41.47 & 42.87 & 1.03 & 0.365 & 10016.7 & 11.2 \\
         & -8 & 41.43 & 42.87 & 1.03 & 0.366 & 10016.7 & 11.2 \\ \midrule
        \multirow{3}{*}{20.20} & -4 & 41.51 & 43.32 & 1.04 & 0.487 & 10123.1 & 11.3 \\
         & -6 & 41.47 & 43.34 & 1.04 & 0.487 & 10126.8 & 11.3 \\
         & -8 & 41.43 & 43.36 & 1.05 & 0.488 & 10131.7 & 11.3 \\ \bottomrule
    \end{tabular}
    \label{tab:radVar}
\end{table}

\subsection{Experimental Facilities}
\label{methods:facilities}

\begin{figure*}[t]
    \centering
    \begin{subfigure}[t]{0.45\textwidth}
        \centering
        \includegraphics[width = 0.9\textwidth]{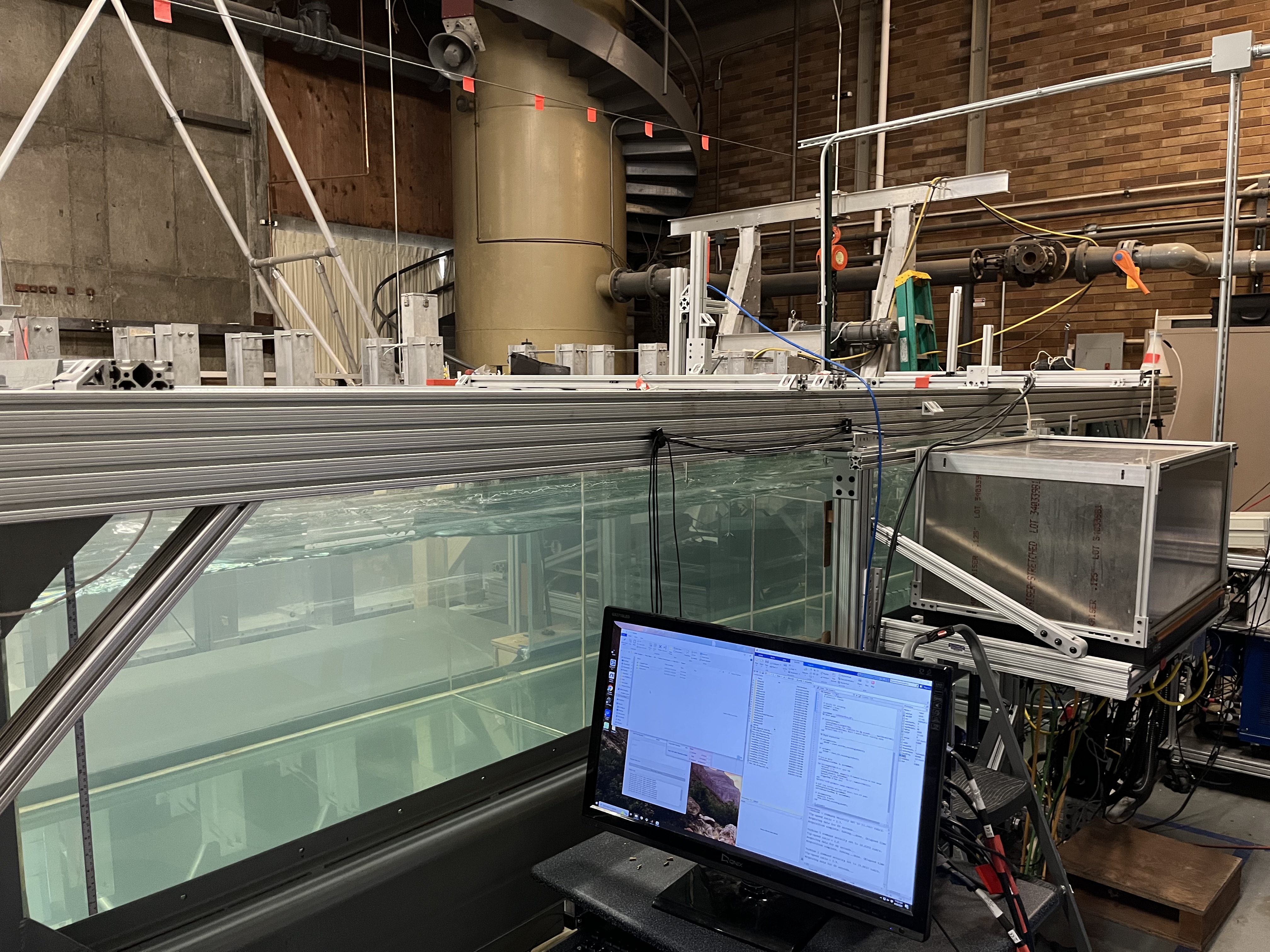}
        \caption{}
        \label{fig:tylerFlume}
    \end{subfigure} \hfill
    \begin{subfigure}[t]{0.45\textwidth}
        \centering
        \includegraphics[width = \textwidth]{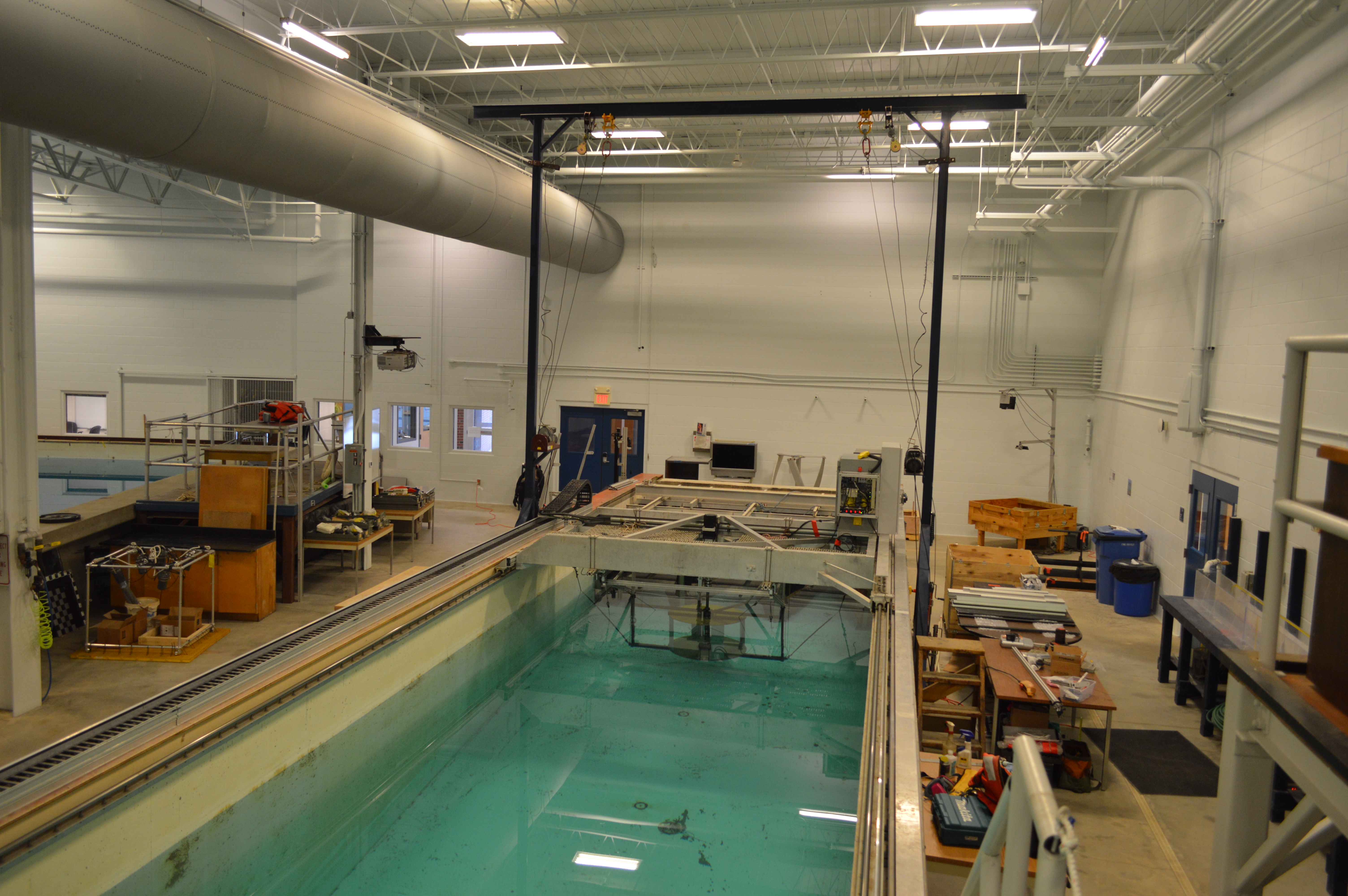}
        \caption{}
        \label{fig:chaseTank}
    \end{subfigure}
    \caption{The experimental facilities used in this study: \subref{fig:tylerFlume} the Tyler flume at UW  and \subref{fig:chaseTank} the towing tank at the Chase Ocean Engineering Laboratory at UNH.}
    \label{fig:expFacilities}
\end{figure*}

Experiments using the Turbine A configurations were conducted in the Alice C. Tyler flume (\Cref{fig:tylerFlume}) at the University of Washington (UW) in Seattle, Washington while experiments using the Turbine B configurations were conducted in the towing tank at the Chase Ocean Engineering Laboratory (\Cref{fig:chaseTank}) at the University of New Hampshire (UNH) in Durham, New Hampshire.

\begin{figure*}[t]
    \begin{subfigure}[t]{0.45\textwidth}
    \centering
        \centering
        \includegraphics[width= 0.7\textwidth]{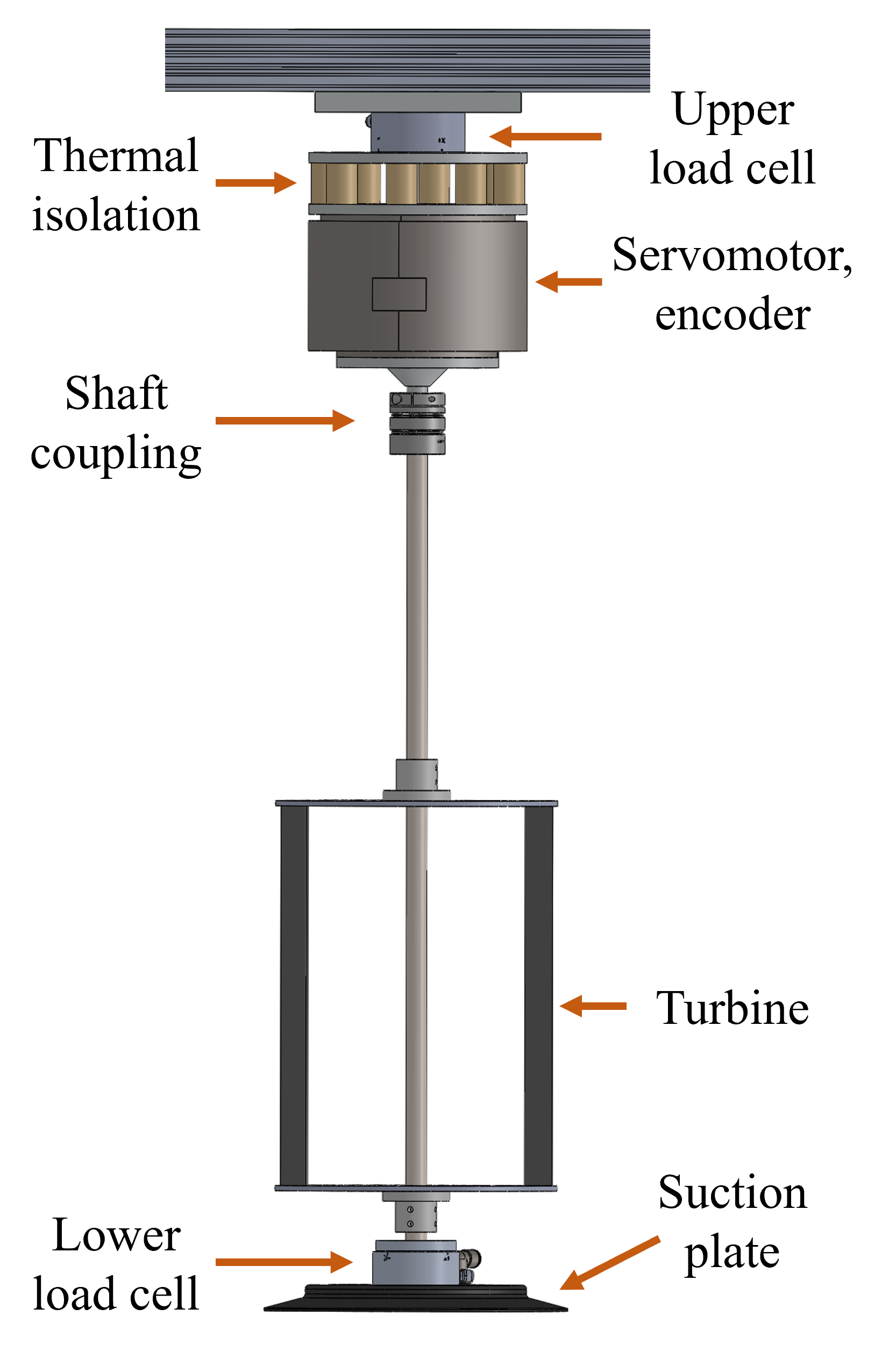}
        \caption{Front view of the experimental setup in the Alice C. Tyler flume with a two-bladed turbine installed.}
        \label{fig:Tyler_setup}
    \end{subfigure} \hfill
    \begin{subfigure}[t]{0.45\textwidth}
        \centering
        \includegraphics[width=\textwidth]{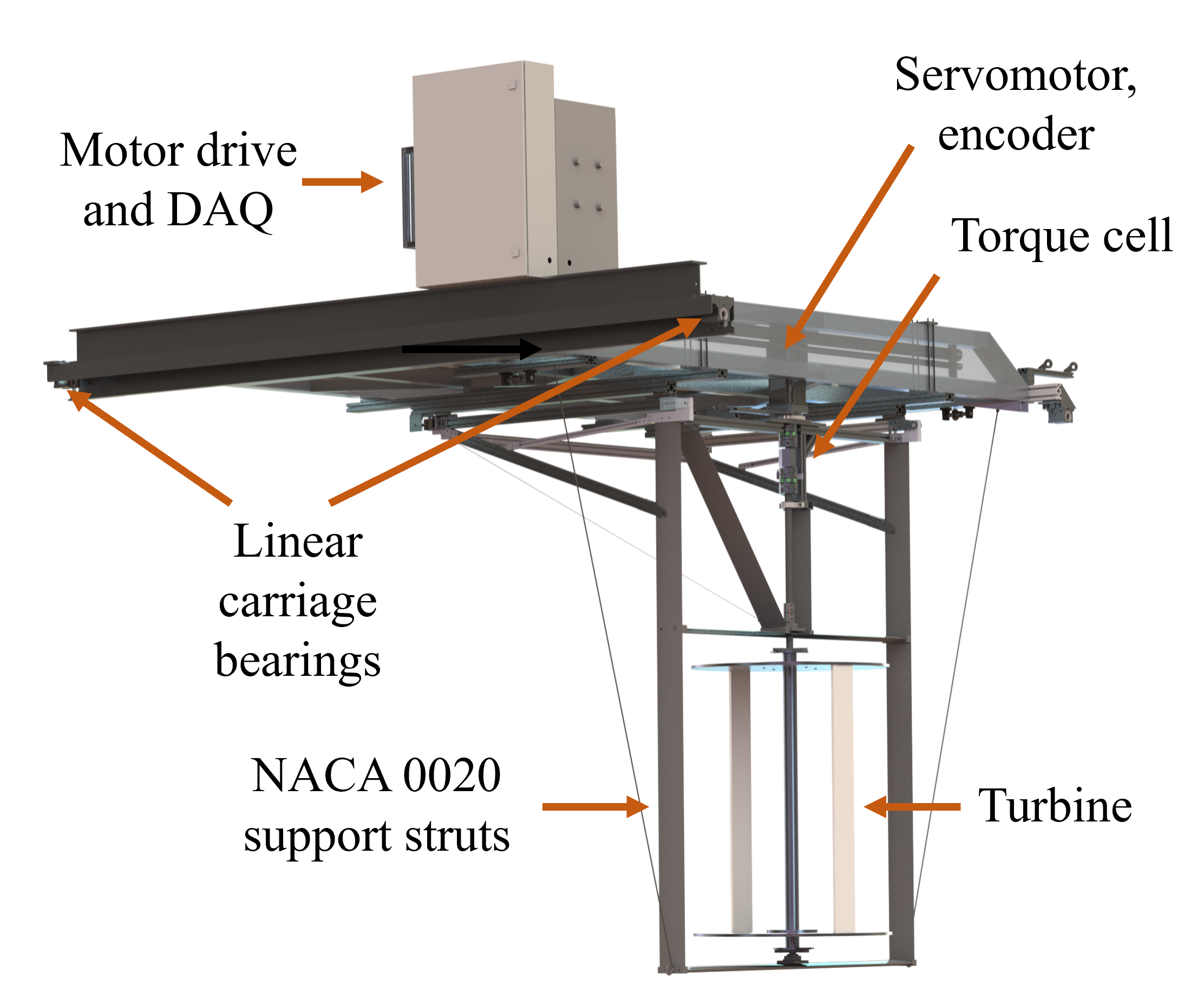}
        \caption{Rendering of the experimental setup used in the Chase Ocean Engineering Laboratory towing tank at UNH. The cross-flow turbine test bed attached to the tow carriage travels to the left during towing.}
        \label{fig:Chase_setup}
    \end{subfigure}
    \caption{The experimental set-ups used in this study.}
    \label{fig:expSetUps}
\end{figure*}

\subsubsection{Alice C. Tyler Flume, University of Washington}
\label{methods:tylerFlume}

The Alice C. Tyler flume is a recirculating free-surface water flume with a test section 0.75 m wide and 4.88 m long. The flow is driven by two pumps with variable frequency drives, which can generate free stream velocities up to 1.1 m/s. Water temperature in the flume is regulated by in-line heating elements and an external chiller. This allows $Re_D$ to be varied by changing both $U_\infty$ and $\nu$, the latter of which is a function of temperature. Due to a large stilling basin, turbulence intensities are relatively low ($2 - 4\%$).

The experimental set-up in the Tyler flume, as shown in \Cref{fig:Tyler_setup}, is similar to that of \citep{strom_impact_2018,polagye_comparison_2019,hunt_effect_2020, ross_effects_2022}. The turbine driveshaft is connected to a servomotor above the water surface (Yaskawa SGMCS-05BC341) via a flexible shaft coupling. The servomotor regulates the turbine's angular velocity and measures the turbine's angular position via an integrated encoder with $2^{16}$ counts per revolution (angular resolution of $0.0055^{\circ}$). The bottom of the turbine driveshaft rests in a bearing secured to the bottom of the flume by a suction plate. Forces and torque acting on the turbine are measured by a pair of six-axis load cells (ATI Mini45-IP65 coupled to the servomotor, ATI Mini40-IP68 coupled to the bottom bearing). When the servomotor regulates the turbine rotation rate to a constant value, the fluid torque acting on the turbine is equal to the sum of the resistive torque imposed on the turbine by the servomotor and the frictional torque from the bottom bearing \citep{polagye_comparison_2019}. Measurements from the servomotor encoder and load cells were streamed to MATLAB (Simulink Desktop Realtime) at 1000 Hz via a DAQ (National Instruments PCIe-6351). The freestream velocity at the turbine midplane was sampled at 64 Hz by an acoustic Doppler velocimeter (ADV) (Nortek Vector) centered laterally in the water column and located five turbine diameters upstream of the axis of rotation. ADV measurements were despiked using the method of Goring and Nikora \citep{goring_despiking_2002}. The freestream water depth was sampled at 5 Hz by an ultrasonic free surface transducer (Omega LVU32) centered laterally in the flume and located two turbine diameters upstream of the ADV. The water temperature was manually monitored throughout experiments (Omega Ultra-Precise RTD) and maintained to within +/- $0.1^{\circ} \mathrm{C}$ of its target value.

For each $c/R$, $\alpha_p$, $N$, and $Re_D$ combination tested, turbine performance was characterized under constant-speed control over a range of $\lambda$ in steps of 0.05 near the performance peak and in steps of 0.1-0.2 further from the peak. Data were collected at each $\lambda$ setpoint for 45 seconds, which was sufficient to obtain convergence for time-average quantities. 

\subsubsection{Chase Ocean Engineering Laboratory, University of New Hampshire}
\label{methods:chaseTank}

The test section of the tow tank at the Chase Ocean Engineering Laboratory measures 3.66 m wide, 2.44 m deep, and 36.6 m long. The experimental setup (\Cref{fig:Chase_setup}) is similar to the smaller-scale version used at UW in that the turbine driveshaft is connected to the test rig at both ends. Complete details of this system are provided by \citet{bachant_physical_2016}. 

The turbine rotation rate is regulated by a servomotor (Kollmorgen AKM62Q) and measured by an integrated encoder with $10^5$ counts per revolution (angular resolution of $0.0036^{\circ}$). Torque is measured by a rotary transducer (Interface T8-200), and losses due to the bottom bearing were characterized by spinning the turbine in air. The turbine test bed is attached to the tow carriage via linear bearings connected to two load cells to measure overall thrust, and the test bed is constructed from extruded aluminum struts (NACA 0020 profile with 0.14 m chord) to minimize drag. The tow carriage speed is measured by a linear encoder (Renishaw LM15). Thrust was measured, but is not reported as subsequent analysis indicated that one of the load cells was malfunctioning. For these experiments, all measured quantities were recorded at 2000 Hz. The temperature of the tow tank was measured manually using a digital thermometer prior to testing.

Each turbine geometry was tested across a range of $\lambda$ in increments of 0.1, with one tow corresponding to one tip-speed ratio. The turbine was spun up to target rotation rate prior to towing. Since the data acquired during each tow include the linear acceleration and deceleration of the test bed, only the measurements when the test bed was at a quasi-steady state are considered.
After towing, the water in the Chase tank was allowed to settle before the next tow began. A saw-toothed geo-textile beach was used to dissipate energy in the tank. The required tank settling times were established by using ADV measurements at mid-tank, mid-height after tows of comparable turbines, from which the time for the measured velocity to drop below the instrument noise floor was determined \citep{bachant_characterising_2015}.
This time was then doubled and used as tank settling time; for the data reported here the settling time between tows was four minutes. 

\subsection{Test Conditions}
\label{methods:flowParams}

\begin{table*}[t]
    \centering
    \caption{Experimental conditions at each test facility.}
        \begin{tabular}{@{}cccccccc@{}}
        \toprule
        \textbf{Test Facility} & \boldmath{$Re_D$} & \boldmath{$U_{\infty}$} \textbf{[m/s]} & \textbf{Temp. [$^{\circ}$C]} & \boldmath{$\nu$} \textbf{[m$^2$/s]} & \boldmath{$h$} \textbf{[m]} & \boldmath{$s$} \textbf{[m]} & \textbf{TI} \textbf{[\%]} \\ \midrule
        Tyler flume              & $ 0.75 \times 10^5$ & 0.60                            & 10.0                           & $1.31 \times 10^{-6}$        & 0.48                 & 0.05                 & 2.5                    \\
        Tyler flume              & $1.6 \times 10^5$  & 0.90                            & 25.0                           & $8.93 \times 10^{-7}$        & 0.48                 & 0.11                 & 2.7                    \\
        Tyler flume              & $2.7 \times 10^5$  & 1.1                            & 38.9                           & $6.72 \times 10^{-7}$        & 0.48                 & 0.17                 & 3.9                    \\
        Chase tank               & $8.3 \times 10^5$  & 1.0                            & 20.3                           & $9.96 \times 10^{-7}$        & 2.44                 & 0.62                 & 0                      \\ \bottomrule
        \end{tabular}
    \label{tab:test_conditions}
\end{table*}

Turbine performance at $Re_D$ = $0.75\times10^5$, $1.6\times10^5$, and $2.7\times10^5$ was evaluated at UW, while performance at $Re_D$ = $8.3\times10^5$ was evaluated at UNH. The velocity-temperature combinations that yield each $Re_D$ (\Cref{eq:ReD}), along with the dynamic water depth ($h$), turbine submergence depth ($s$), and turbulence intensity (TI) at each test condition, are summarized in \Cref{tab:test_conditions}. It is noted that because multiple $c/R$ were tested at each $Re_D$, the chord-based Reynolds number, $Re_c$ (\Cref{eq:ReC}), was simultaneously varied during these experiments. Although this could be counteracted by changing $\nu$ accordingly at each $c/R$, this would restrict the testable range of $Re_D$. Additionally, as $U_{\mathrm{rel}}$ cannot be measured, the actual $Re_c$ for each turbine would still remain indeterminate. The implications of these $Re_c$ variations at constant $Re_D$ across the tested geometric parameter space are further discussed in \Cref{disc:reynolds}.

In addition to the Reynolds number, other non-dimensional flow parameters impact turbine performance. One such parameter is the channel blockage \citep{garrett_efficiency_2007,bahaj_power_2007,consul_blockage_2013,schluntz_effect_2015,houlsby_power_2017,ross_experimental_2020,ross_effects_2022}, which is defined as the ratio of the turbine rotor's projected area to the channel cross-sectional area,
\begin{equation}
    \beta = \frac{2HR'}{hw},
    \label{eq:blockage}
\end{equation}
\noindent where $w$ is the channel width. The size of the turbine used in each facility resulted in a constant nominal blockage ratio of $\approx\!11\%$ across all tests (\Cref{tab:radVar}). The implications of blockage effects for these experiments are discussed in \Cref{disc:blockage}.

In addition to the blockage ratio and Reynolds number, the Froude numbers based on channel depth, $h$ ($Fr_h = U_\infty / \sqrt{gh}$) \citep{consul_blockage_2013,vogel_effect_2016,gauvin-tremblay_twoway_2020,hunt_effect_2020}, and turbine submergence depth, $s$ ($Fr_s = U_\infty / \sqrt{gs}$) \citep{birjandi_power_2013,kolekar_performance_2015,ross_effects_2022}, have both been shown to affect turbine performance, although to a much lesser degree than either the Reynolds number or blockage \citep{ross_effects_2022}. At UW, as $U_\infty$ was varied for each $Re_D$, the submergence depth of the turbine rotor was varied to hold $Fr_s$ approximately constant at 0.85. At UNH, $Fr_s$ was approximately 0.41. Across both facilities, $Fr_h \leq 0.51$, such that all tests were performed in subcritical regimes.

\subsection{Dimensionless Performance}
\label{methods:perfMetrics}

The mechanical efficiency of the turbine is quantified using the power coefficient, $C_P$, which is the ratio of the mechanical power produced by the turbine to the kinetic power passing through the rotor projected area,
\begin{equation}
    \label{eq:cp}
    C_P(t) =  \frac{Q(t)\omega(t)}{\frac{1}{2} \rho A U^3_\infty(t)} \ \ ,
\end{equation}

\noindent where $Q(t)$ is the instantaneous fluid torque on the turbine as shown in \Cref{fig:turbOverhead}, $\omega(t)$ is the instantaneous angular velocity, $\rho$ is the fluid density (calculated from the water temperature), $A = 2HR'$ is the rotor projected area, and $U_{\infty}(t)$ is the instantaneous freestream velocity.
$C_P$ includes losses due to parasitic torque on the disk end plates which are relatively high \citep{strom_impact_2018} and can obscure blade-level trends. For example, as the tip-speed ratio increases, gains in blade-level power generation can be more than offset by parasitic losses from the end plates; this complicates the comparison of performance across turbine geometries with different optimal tip-speed ratios. To compensate for this, the efficiency of the turbine blades alone is estimated via a linear superposition principle in which the losses associated with the support structures (dominated by the end plates) are subtracted from measurements of turbine power at the same rotation rate \citep{bachant_experimental_2016,strom_impact_2018}. In other words,
\begin{equation}
    C_{P,\mathrm{blades}}(Re_D,\lambda) \approx C_{P,\mathrm{turbine}}(Re_D,\lambda) - C_{P,\mathrm{supports}}(Re_D,\lambda).
    \label{eq:cpb}
\end{equation}

The streamwise forces on the turbine are characterized using the thrust coefficient, defined as
\begin{equation}
    C_T(t) = \frac{T(t)}{\frac{1}{2} \rho A U^2_\infty(t)}\ \ ,
    \label{eq:ct}
\end{equation}
\noindent where $T(t)$ is the instantaneous force on the turbine in the direction of the freestream as shown in \Cref{fig:turbOverhead}. Unlike for $C_P$, blade-level $C_T$ cannot be estimated by subtracting the $C_T$ of a bladeless turbine. Whereas the parasitic torque associated with rotating the drive shaft is small in comparison to the parasitic torque imposed by the end plates, the thrust force on the drive shaft is appreciable and biased when the induction associated with energy harvesting is absent. Consequently, $C_T$ for the full turbine is presented here.
Similarly, the cross-stream loading on the turbine can be characterized via the lateral force coefficient, defined as
\begin{equation}
    C_L(t) = \frac{L(t)}{\frac{1}{2} \rho A U^2_\infty(t) }\ \ ,
    \label{eq:cl}
\end{equation}
\noindent where $L(t)$ is the instantaneous force on the turbine in the horizontal direction perpendicular to the freestream velocity as shown in \Cref{fig:turbOverhead}. As neither the end plates nor the central shaft produce appreciable lateral force, the turbine and blade-only values of $C_L$ are approximately equivalent.

At UW, measurements of $Q(t)$, $T(t)$, $L(t)$, and $\omega(t)$ are synchronized, but are asynchronous with measurements of $U_{\infty}(t)$. Therefore, the time-averages of $U^2_{\infty}(t)$ and $U^3_{\infty}(t)$ at a given $\lambda$ set-point are used to calculate $C_P(t)$, $C_T(t)$, and $C_L(t)$. This approximation is a reasonable approach for experimental conditions with low turbulence intensity. While empirical corrections for advection time between the velocity measurement and turbine can be applied (e.g.,  \citep{polagye_comparison_2019}), this can increase the range of observed $C_P(t)$ relative to power if Taylor's frozen field hypothesis is violated. 
At UNH, measurements of $Q(t)$, $\omega(t)$, and $U_{\infty}(t)$ (obtained from the tow tank linear encoder) are synchronized, but the same method was used to calculate $C_P(t)$ since the variations in tow velocity were even lower than that of the inflow velocity in the Tyler flume.
For both facilities, an integer number of turbine rotations are used to calculate turbine performance parameters at each $\lambda$.
Experimental uncertainty for $C_P$ measurements by both systems is discussed in \Cref{appendix:repeatability}. 

\subsection{Synthesis}

\begin{figure*}[t]
    \centering
    \includegraphics[width = \textwidth]{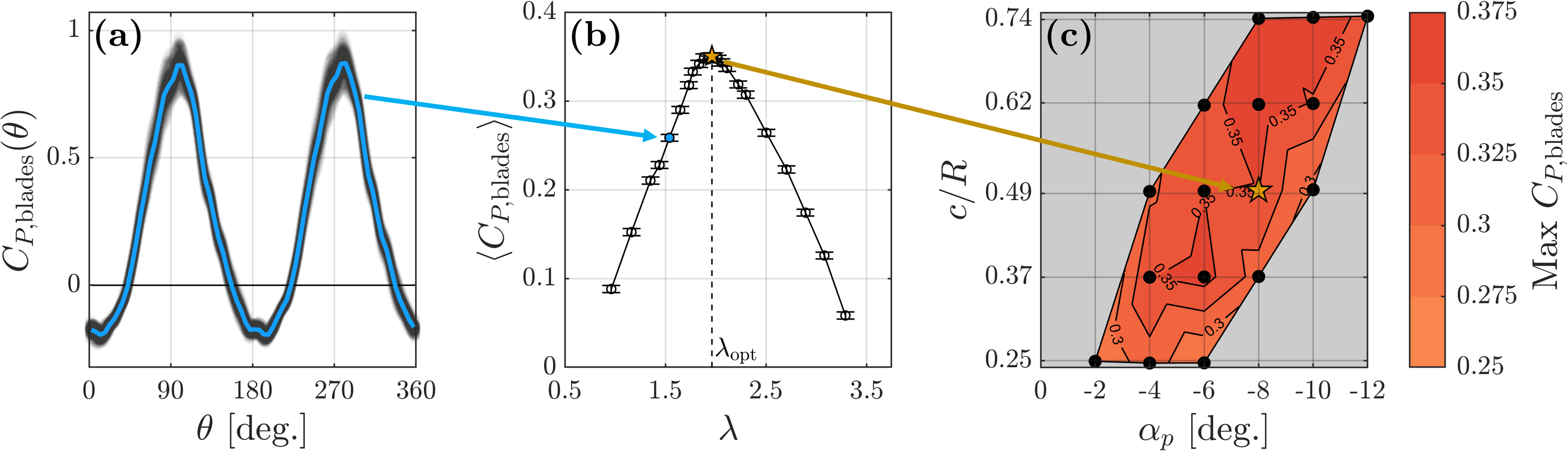}

    {\phantomsubcaption\label{fig:phaseAveSynth}
    \phantomsubcaption\label{fig:timeAveSynth}
    \phantomsubcaption\label{fig:heatmapSynth}
    }
    \caption{Construction of performance contours from instantaneous $C_{P,\mathrm{blades}}$ as a function of $c/R$ and $\alpha_p$ at $Re_D = 1.6 \times 10^5$ and $N = 2$. \subref{fig:phaseAveSynth} Instantaneous blade-level performance during a cycle at a given $\lambda$. The blue curve indicates the median cycle. \subref{fig:timeAveSynth} Time-average $C_{P,\mathrm{blades}}-\lambda$ curve, with performance peak and corresponding $\lambda$ annotated. \subref{fig:heatmapSynth} Placement of the maximum performance point for this turbine on a contour map for $Re_D = 1.6 \times 10^5$ and $N = 2$.}
    \label{fig:methods_three_panel}
\end{figure*}

To visualize trends from experimental data, results are aggregated and presented as contour maps of turbine performance metrics (e.g., maximum $C_{P,\mathrm{blades}}$) as a function of $c/R$, $\alpha_p$, $N$, and $Re_D$.
\Cref{fig:methods_three_panel} graphically demonstrates the contour map generation process for maximum $C_{P,\mathrm{blades}}$ at one $Re_D$ and $N$ from the collected data. For a given $c/R$ and $\alpha_p$, turbine performance was characterized under constant speed control over a range of $\lambda$. As shown in \Cref{fig:phaseAveSynth}, during each cycle at a given $\lambda$, $C_{P,\mathrm{blades}}(\theta)$ oscillates as a consequence of the periodic fluid dynamics and, for some cases, switches between power production and power consumption within a rotational cycle. 
In addition to the phase-varying oscillations in $C_P$, there are cycle-to-cycle variations in the performance at each $\theta$, primarily as a consequence of fluctuations in the inflow conditions \citep{snortland_cycletocycle_2023}. Time-average $C_{P,\mathrm{blades}}$ for this operating condition is represented as a point on a characteristic performance curve (\Cref{fig:timeAveSynth}). The operating condition associated with the maximum $C_{P,\mathrm{blades}}$ for this geometry is designated as $\lambda_{\mathrm{opt}}$, and this maximum performance is represented as a single point on a map of $c/R$ versus $\alpha_p$ for one $Re_D\!-\!N$ combination (\Cref{fig:heatmapSynth}). This process is repeated for each $c/R\!-\!\alpha_p$ combination tested at that $Re_D\!-\!N$ combination, and linear interpolation is used to visualize the $C_{P,\mathrm{blades}}$ contours across all combinations of $c/R$ and $\alpha_p$ within the convex hull spanned by the tested configurations.  In \Cref{sec:results}, $C_{P,\mathrm{blades}}$ contours at each $Re$ and $N$ are tessellated to visualize the entire parameter space. Similar techniques are used to visualize $\lambda_{\mathrm{opt}}$, $C_T$, and $C_L$ associated with maximum $C_{P,\mathrm{blades}}$, as well as the standard deviation of the phase-median performance for these metrics at $\lambda_{\mathrm{opt}}$ (i.e., variation of metrics within a rotational cycle).


\section{Results}
\label{sec:results}

\begin{figure*}[t]
    \centering
    \includegraphics[width = \textwidth]{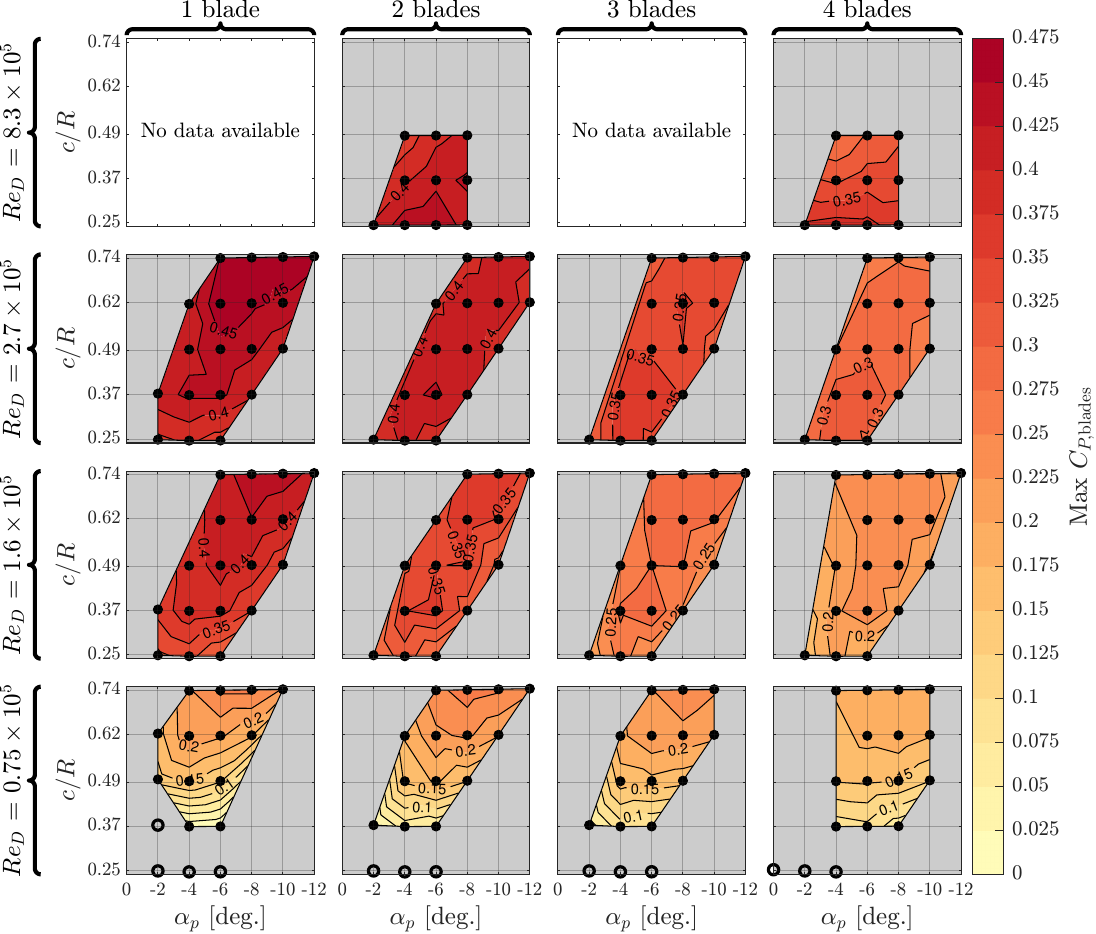}
    \caption{Maximum blade-level $C_P$ contours as a function of $c/R$ and $\alpha_p$ at each $Re_D$ and $N$ tested. Points with open circles indicate geometries that did not produce power at any tip-speed ratio and, as such, have no optimal operating condition.}
    \label{fig:cpBladeHeatmap}
\end{figure*}

This section includes aggregate results of all experiments. While general trends are highlighted, a more comprehensive discussion of the fluid dynamics that underlie these trends is deferred to \Cref{sec:discussion}.

\Cref{fig:cpBladeHeatmap} shows maximum blade-level efficiency, $C_{P,\mathrm{blades}}$, across the tested parameter space, from which several trends emerge. First, for a given turbine geometry, as $Re_D$ increases (i.e., moving bottom-to-top along one column of \Cref{fig:cpBladeHeatmap}), maximum $C_{P,\mathrm{blades}}$ increases, which is in agreement with prior work \Citep{bachant_effects_2016,bachant_experimental_2016,miller_verticalaxis_2018, miller_solidity_2021, ross_effects_2022}. While the maximum $C_{P,\mathrm{blades}}$ increases with $Re_D$ for all tested geometries, the rate at which maximum $C_{P,\mathrm{blades}}$ increases varies with geometry. Reynolds independence of maximum $C_{P,\mathrm{blades}}$ is not explicitly observed, which is to be expected given the relatively low Reynolds numbers involved. However, based on previous experiments in the Chase towing tank with turbines of similar size and $Re_D$ (e.g., \cite{bachant_effects_2016, bachant_experimental_2016}), the experiments reported here were likely approaching Reynolds independence. 

Second, as the number of blades increases (i.e., moving left-to-right along one row of \Cref{fig:cpBladeHeatmap}), the maximum $C_{P,\mathrm{blades}}$ decreases. Decreasing maximum efficiency with increasing blade count is well documented in prior work \citep{mcadam_experimental_2013, li_effect_2015, araya_transition_2017, miller_solidity_2021}, and this relationship holds for nearly all of the tested turbine geometries, with the exception of some of the lowest $c/R$ and $Re_D$ tested.

\begin{figure*}[t]
    \centering
    \includegraphics[width = \textwidth]{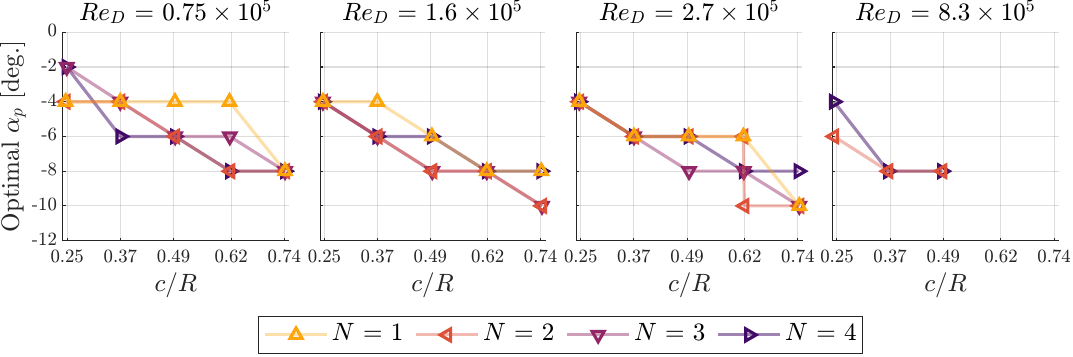}
    \caption{Optimal preset pitch angle for each $c/R$ across all $N$ and $Re_D$ tested.}
    \label{fig:pitch_CToR_Lineplot}
\end{figure*}

Third, at a given $Re_D$ and $N$ (i.e., within a single tile of \Cref{fig:cpBladeHeatmap}), as $c/R$ increases, the $\alpha_p$ for maximum performance becomes more negative. In other words, blades with higher chord-to-radius ratios are more efficient with greater toe-out preset pitch. This relationship is highlighted in \Cref{fig:pitch_CToR_Lineplot}. While first identified in early, narrower work by \citet{takamatsu_experimental_1991}, this trend has not previously been systematically demonstrated across a wide operational and geometric range. Additionally, the optimal $\alpha_{p}$ for each $c/R$ decreases slightly with $Re_D$.
Although this trend is weaker, it is supported by a similar observation by \citet{somoano_dead_2018}.

\begin{figure}[t]
    \centering
    \includegraphics[width = \textwidth]{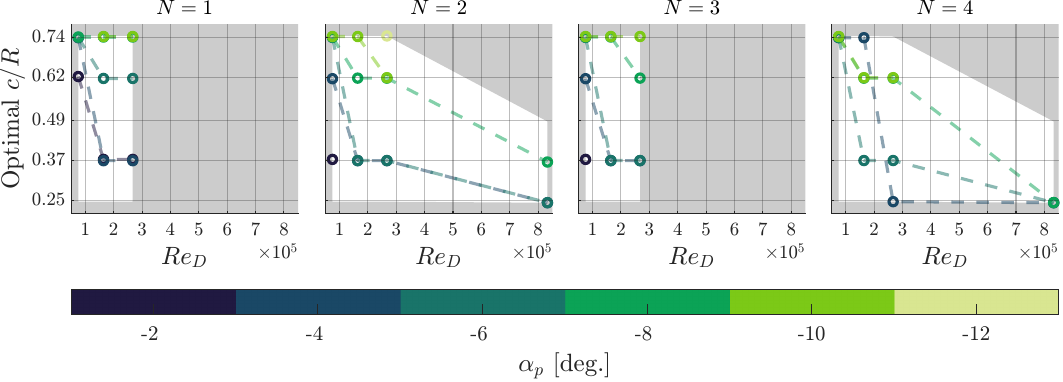}
    \caption{Optimal $c/R$ as a function of $Re_D$ for each $\alpha_p$ and $N$. The grey shaded region denotes the $c/R$-$Re_D$ domain outside of the parameter space explored.}
    \label{fig:cToR_Re_Lineplot}
\end{figure}

Fourth, for a given $\alpha_{p}$, optimal $c/R$ decreases with $Re_D$. This trend is highlighted in \Cref{fig:cToR_Re_Lineplot}, and is especially clear for the $N = 2$ and $N = 4$ cases, which span the largest range of $Re_D$. Though this trend is implicitly apparent through comparison of separate prior studies, this is the first time it is shown in a single set of experiments with all other parameters controlled. It is also notable that, for several of the geometries evaluated in this work, the $c/R$ that yields maximum performance at a given $Re_D$ was the minimum or maximum $c/R$ tested. Therefore, for these rotors, it is possible that the true optimal $c/R$ lies outside of the tested domain (i.e., in the gray shaded region in \Cref{fig:cToR_Re_Lineplot}).
Additionally, at all $Re_D$, as $N$ increases, the performance of low $c/R$ blades approaches ---and at some $Re_D$, surpasses---that of high $c/R$ blades (\Cref{fig:cpBladeHeatmap}). In other words, at high $N$, blades with low $c/R$ become more efficient, especially at higher $Re_D$.

\begin{figure*}[t]
    \centering
    \includegraphics[width=\textwidth]{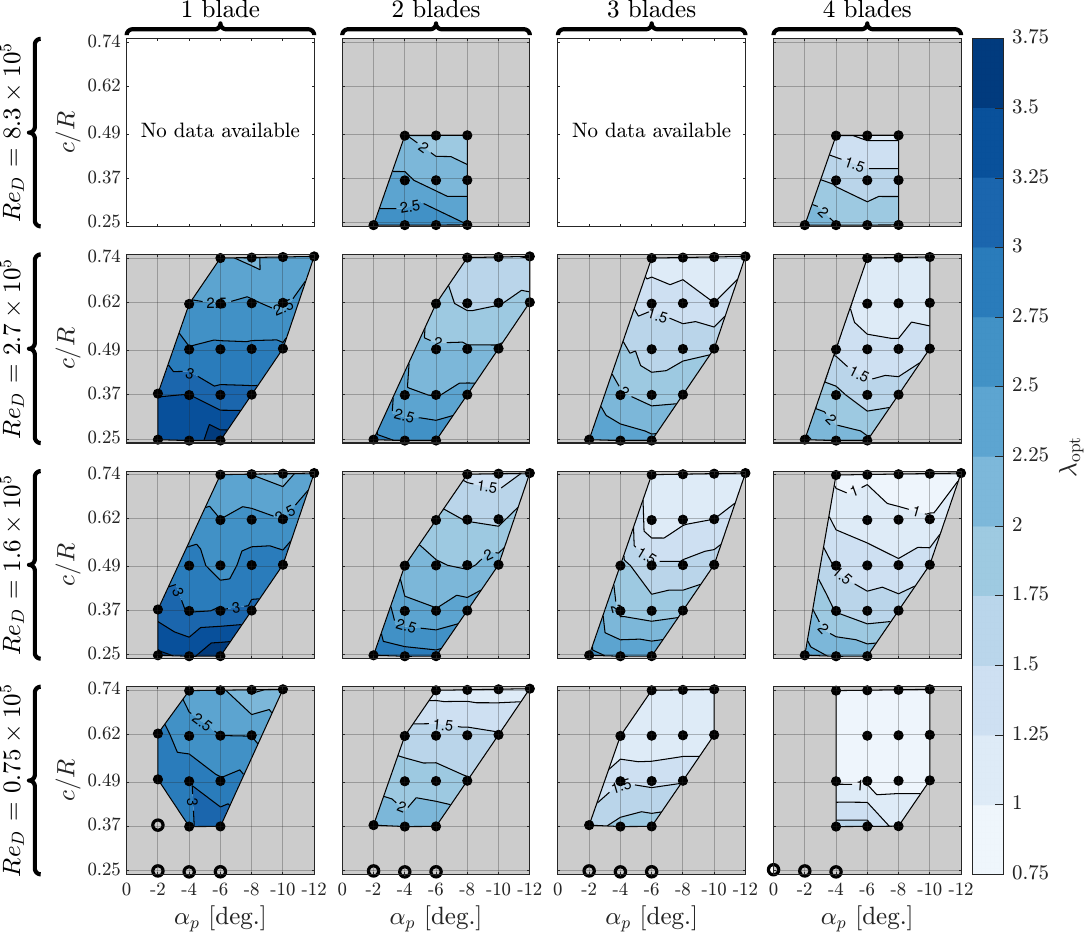}
    \caption{Contours of $\lambda_{\mathrm{opt}}$ (based on $C_{P,\mathrm{blades}}$) as a function of $c/R$ and $\alpha_p$ at each $Re_D$ and $N$ tested. Points with open circles indicate geometries that did not produce power at any tip-speed ratio and, as such, have no optimal operating condition.}
    \label{fig:tsrHeatmap}
\end{figure*}

Contours of $\lambda_{\mathrm{opt}}$ across the parameter space are shown in \Cref{fig:tsrHeatmap}. For all $Re_D$ and $N$ tested, the optimal tip-speed ratio decreases as $c/R$ increases. Additionally, the optimal tip-speed ratio decreases as $N$ increases. Unlike $C_{P,\mathrm{blades}}$, $\lambda_{\mathrm{opt}}$ is relatively independent of $Re_D$ and $\alpha_{p}$ in the range tested, suggesting that the optimal tip-speed ratio may be invariant with turbine scale, at least above some threshold value for $Re_D$ or within the range of $\alpha_p$ tested at each $c/R$.


\begin{figure}[t]
    \centering
    \includegraphics[width=\textwidth]{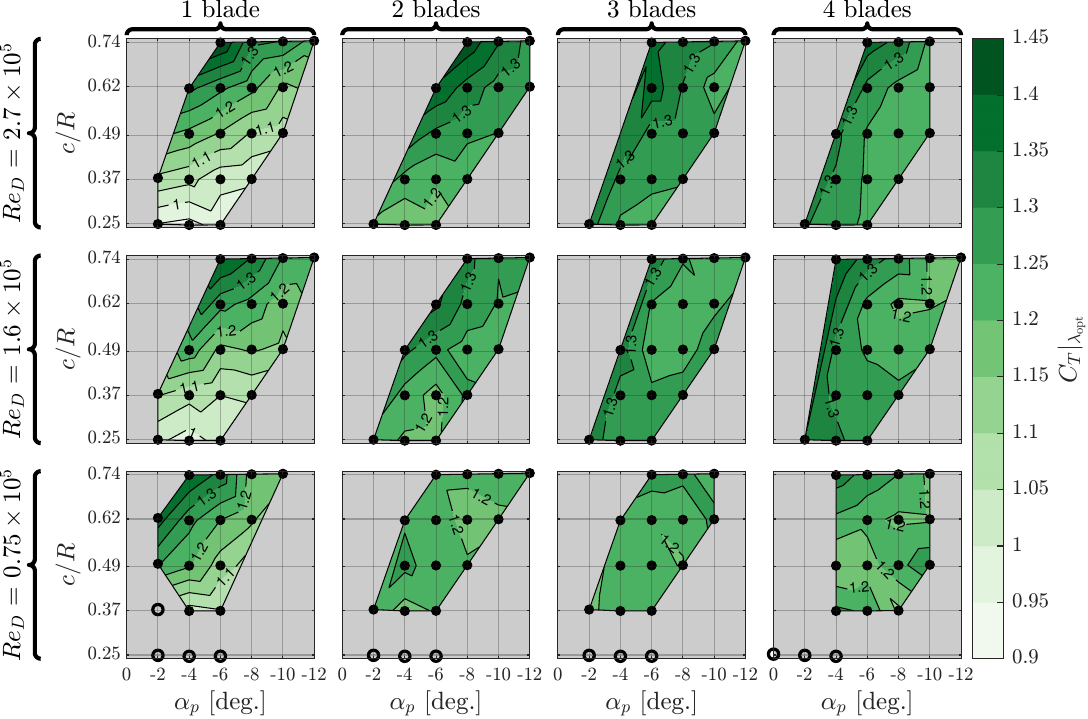}
    \caption{Contours of time-average $C_T$ at $\lambda_{\mathrm{opt}}$ for blade performance as a function of $c/R$ and $\alpha_p$ at each $Re_D$ and $N$ tested. Points with open circles indicate geometries that did not produce power at any tip-speed ratio and, as such, have no optimal operating condition.}
    \label{fig:ctBladeOptHeatmap}
\end{figure}

\begin{figure*}[t]
    \centering
    \includegraphics[width=\textwidth]{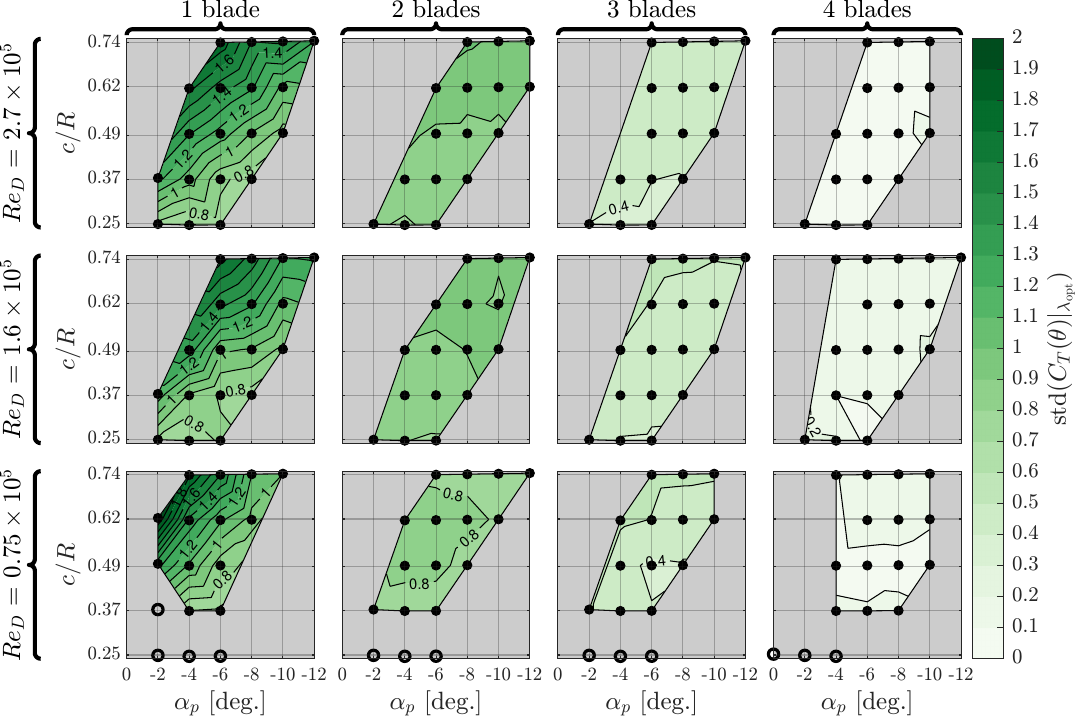}
    \caption{Contours of the standard deviation of $C_{T}(\theta)$ at $\lambda_{\mathrm{opt}}$ as a function of $c/R$ and $\alpha_p$ at each $Re_D$ and $N$ tested. Points with open circles indicate geometries that did not produce power at any tip-speed ratio and, as such, have no optimal operating condition.}
    \label{fig:ctVarHeatmap}
\end{figure*}

\Cref{fig:ctBladeOptHeatmap} shows contours of the time-average thrust force coefficient at $\lambda_{\mathrm{opt}}$ (i.e., $C_T\rvert_{\lambda_{\mathrm{opt}}}$) across the parameter space. As noted in \Cref{methods:perfMetrics}, the contours shown are of full turbine $C_T$, as opposed to blade-level $C_T$. Consequently, trends in actual blade-level $C_T$ may differ somewhat since the measured full-turbine $C_T$ values include thrust force on the central shaft and end plates.
Compared to the differences in maximum $C_{P,\mathrm{blades}}$ across the parameter space in \Cref{fig:cpBladeHeatmap}, differences in $C_T\rvert_{\lambda_{\mathrm{opt}}}$ are less significant. For $N = 1$, $C_T\rvert_{\lambda_{\mathrm{opt}}}$ increases with both $c/R$ and $\alpha_p$. However, for $N > 1$, trends in $C_T\rvert_{\lambda_{\mathrm{opt}}}$ across the parameter space are more ambiguous.
While prior research has noted slight increases in $C_T\rvert_{\lambda_{\mathrm{opt}}}$ with $Re_D$ \citep{bachant_effects_2016} and $N$ \citep{li_effect_2015}, neither trend is observed in the present data. However, because few studies report $C_T$ measurements, further comparison to prior work is limited.
In contrast, periodic variation of the thrust coefficient within a single rotational cycle (e.g., $C_T(\theta)$) is strongly influenced by turbine geometry. \Cref{fig:ctVarHeatmap} shows contours of the standard deviation of $C_{T}(\theta)\rvert_{\lambda_{\mathrm{opt}}}$ for each rotor geometry. For single-bladed turbines, variability in $C_{T}(\theta)\rvert_{\lambda_{\mathrm{opt}}}$ is highest, and tends to increase as $c/R$ increases and $\alpha_{p}$ becomes more positive. The variability in $C_{T}(\theta)\rvert_{\lambda_{\mathrm{opt}}}$ decreases with each blade added, and, for $N > 1$, neither $c/R$ nor $\alpha_{p}$ substantially effects the the variation of $C_{T}(\theta)\rvert_{\lambda_{\mathrm{opt}}}$.
Notably, $Re_D$ has little effect on both $C_T\rvert_{\lambda_{\mathrm{opt}}}$ and the standard deviation of $C_T(\theta)\rvert_{\lambda_{\mathrm{opt}}}$ in the range tested. Similar trends are observed for the standard deviations of $C_{P,\mathrm{blades}}(\theta)\rvert_{\lambda_{\mathrm{opt}}}$ and $C_{L}(\theta)\rvert_{\lambda_{\mathrm{opt}}}$ across the tested parameter space.

\begin{figure}[t]
    \centering
    \includegraphics[width=\textwidth]{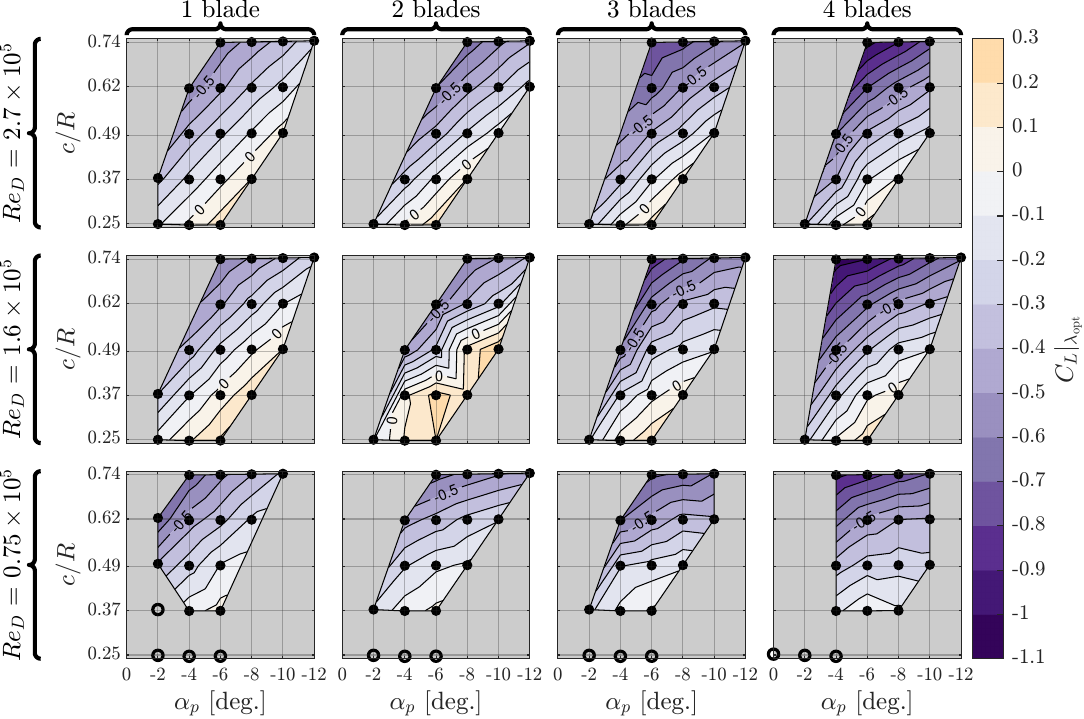}
    \caption{Contours of time-average $C_L$ at $\lambda_{opt}$ for blade performance as a function of $c/R$ and $\alpha_p$ at each $Re_D$ and $N$ tested. Points with open circles indicate geometries that did not produce power at any tip-speed ratio and, as such, have no optimal operating condition.}
    \label{fig:clBladeOptHeatmap}
\end{figure}

\Cref{fig:clBladeOptHeatmap} shows contours of the time-average lateral force coefficient at $\lambda_{\mathrm{opt}}$ (i.e., $C_{L}\rvert_{\lambda_{\mathrm{opt}}}$) across the parameter space. While the magnitude of $C_{L}\rvert_{\lambda_{\mathrm{opt}}}$ for each geometry is less than that of $C_{T}\rvert_{\lambda_{\mathrm{opt}}}$, the range of $C_{L}\rvert_{\lambda_{\mathrm{opt}}}$ across the tested geometries is larger. Across all $Re_D$ and $N$ tested, the magnitude of $C_{L}\rvert_{\lambda_{\mathrm{opt}}}$ increases most strongly with increasing $c/R$ and more positive $\alpha_p$. While decreasing $\alpha_p$ at a given $c/R$ can reduce the magnitude of the time-average $C_{L}\rvert_{\lambda_{\mathrm{opt}}}$, the contours of \Cref{fig:clBladeOptHeatmap} imply that beyond the range tested, further decreases in $\alpha_p$ will reverse the direction of the time-average $C_{L}\rvert_{\lambda_{\mathrm{opt}}}$ and eventually increase its magnitude.
However, because the underlying fluid mechanics of $C_L$ are difficult to interpret without knowledge of the forces acting on individual blades, further discussion of the experimental results is primarily focused on $C_{P,\mathrm{blades}}$ and $C_T$.

\section{Discussion}
\label{sec:discussion}

To interpret the experimental results given in \Cref{sec:results}, the fluid dynamic context for the observed trends is first established.
Then, considering the experimental results and prior research, the effects of each geometric parameter are analyzed.
Finally, this section concludes with generalized design considerations for cross-flow turbine rotors.

\subsection{Contextual Fluid Dynamics}

\subsubsection{Turbine-Level vs. Blade-Level Performance}
\label{disc:turbVBlade}

\begin{figure}[t]
    \centering
    \includegraphics[width=0.5\linewidth]{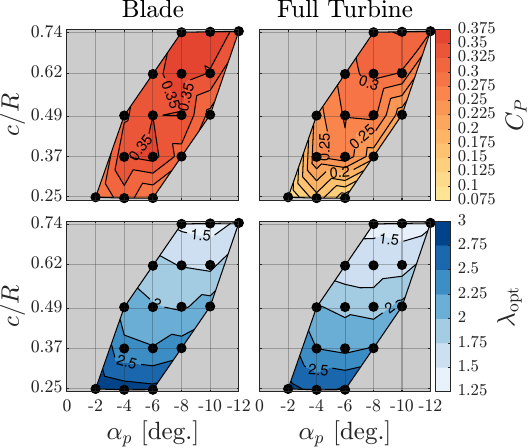}
    \caption{Contours of maximum $C_{P,\mathrm{blades}}$ and corresponding $\lambda_{\mathrm{opt}}$ (left column) compared to $C_P$ and corresponding $\lambda_{\mathrm{opt}}$ for the full turbine (right column) at $N=2$ and $Re_D = 1.6\times10^5$.}
    \label{fig:turbVBlade}
\end{figure}

In \Cref{sec:results}, $C_{P,\mathrm{blades}}$ is presented rather than full turbine $C_P$, which includes the effects of the support structures for mounting the blades (\Cref{eq:cpb}).
In practice, support structures are obviously required, and for full-scale deployments their design is informed by both structural constraints and turbine operating condition.
For relatively low $\lambda$, the parasitic torque from the support structures is limited and the choice of support structure can be unimportant.
As $\lambda$ increases, the parasitic losses from the support structures can become considerable and an important part of the design process \citep{bianchini_design_2015}.

\Cref{fig:turbVBlade} illustrates how the inclusion of these losses alters the contours of maximum $C_{P,\mathrm{blades}}$ and $\lambda_{\mathrm{opt}}$ for turbines at $Re_D = 1.6\times10^5$ and $N=2$.
Relative to $C_{P,\mathrm{blades}}$, the contours of full turbine $C_P$ at this $Re_D\!-\!N$ combination show greater dependence on $c/R$, with the largest tested $c/R$ significantly outperforming the lowest tested $c/R$.
However, this is, in part, because blades with lower $c/R$ perform best at higher rotation rates, and consequently incur greater support structure losses at peak blade performance.
When losses from the support structures are subtracted (\Cref{eq:cpb}), blades with low $c/R$ are not penalized for operating at higher $\lambda_{\mathrm{opt}}$, and their performance relative to large $c/R$ blades is improved.

While this demonstrates the importance of considering support structure losses, discussion of the experimental results focuses on $C_{P,\mathrm{blades}}$ rather than full turbine $C_P$ to identify trends that are independent of support structure and facilitate the broader applicability of this data set. As support structure losses depend not only on rotation rate, but also the shape (e.g., rectangular strut, foil strut, end plate, etc.) and mounting location (e.g., mid-span or blade-end) of the support \citep{bianchini_design_2015, bachant_experimental_2016, strom_impact_2018, villeneuve_increasing_2021}, trends in full turbine $C_P$ in these experiments would likely change if different supports were used.
Additionally, numerical studies of cross-flow turbines often omit support structures to reduce computational expense, such that $C_{P,\mathrm{blades}}$ is most relevant for the validation of these studies.

\subsubsection{Blockage Effects}
\label{disc:blockage}

All experiments involved turbines operating in moderately confined flow ($\approx\!11\%$ blockage ratio) and the observed $C_{P,\mathrm{blades}}$, $C_T$, $C_L$, and $\lambda_{\mathrm{opt}}$ are consequently elevated relative to unconfined conditions \citep{garrett_efficiency_2007, consul_influence_2009, houlsby_power_2017, ross_experimental_2020}. 
However, the actual blockage effects experienced by each turbine depend on its geometry and operating condition. Setting aside minor differences in blockage associated with projected area (\Cref{tab:radVar}), blockage effects are a function of both the channel blockage ratio (\Cref{eq:blockage}) and the rotor thrust coefficient \citep{garrett_efficiency_2007, houlsby_power_2017, ross_experimental_2020}. 
Consequently, even if two turbines have the same blockage ratio, the turbine with the higher $C_T$ will experience greater blockage-driven augmentations to $C_{P,\mathrm{blades}}$. Since $C_{T}\rvert_{\lambda_{\mathrm{opt}}}$ varies by $\approx\!45\%$ within the tested parameter space (\Cref{fig:ctBladeOptHeatmap}), the influence of $C_T$ on blockage effects is non-negligible. However, the largest variation in $C_{T}\rvert_{\lambda_{\mathrm{opt}}}$ occurs for single-bladed bladed turbines, with variations $<$ 25\% for all turbines with $N > 1$. Therefore, while blockage effects were non-uniform across the parameter space, the distortion of the results by blockage effects is likely limited for $N > 1$.

To compensate for blockage effects, it is possible to apply analytical corrections to experimental data and estimate performance in unconfined flow \citep{ross_experimental_2020}. However, a blockage correction has not been applied to these data for several reasons. 
First, because no thrust data were collected at UNH, corrections could not be applied to performance measurements at all $Re_D$. 
Second, it is unclear whether blockage corrections are directly applicable to $C_{P,\mathrm{blades}}$ or if this would require a reliable estimate of the blade-level thrust coefficient.
Finally, since blockage corrections are typically applied to time-averaged performance, a focus on blockage-corrected results would preclude discussion of how rotor geometry influences periodic variability in $C_{P,\mathrm{blades}}$ and $C_T$ as a function of azimuthal position.

\subsubsection{Reynolds Number Effects}
\label{disc:reynolds}

Turbine performance is affected by the local blade Reynolds number, $Re_c$ (\Cref{eq:ReC}). As $Re_c$ increases, the boundary layer on the blade surface becomes more turbulent and resistant to the adverse pressure gradients that cause flow separation on the suction side of the blade at higher angles of attack \citep{lissaman_lowreynoldsnumber_1983}
Since this improves the blade's lift-to-drag ratio for a given kinematic condition, efficiency increases with $Re_c$ until an independence threshold is reached \citep{bachant_effects_2016, miller_verticalaxis_2018,miller_solidity_2021}.

However, the results presented in \Cref{sec:results} are categorized by the diameter-based Reynolds number, $Re_D$; this definition is used for expediency ($D$, $U_{\infty}$, and $\nu$ are easily measured) and because it is agnostic to blade geometry and $\lambda$, which are both intentionally varied across the parameter space.
If induction is neglected, the nominal $Re_c$ can be expressed in terms of other experimental parameters by combining equations \Cref{eq:ReC,eq:ReD,eq:U_n}, yielding
\begin{equation}
    Re_{c,n} = \frac{U_{\mathrm{rel},n} c}{\nu} = Re_D \frac{c}{R} \frac{\sqrt{\lambda^2 + 2\lambda\cos{\theta} + 1}}{2}.
    \label{eq:reRelation}
\end{equation}

\begin{figure*}[t]
    \centering
    \includegraphics[width=\textwidth]{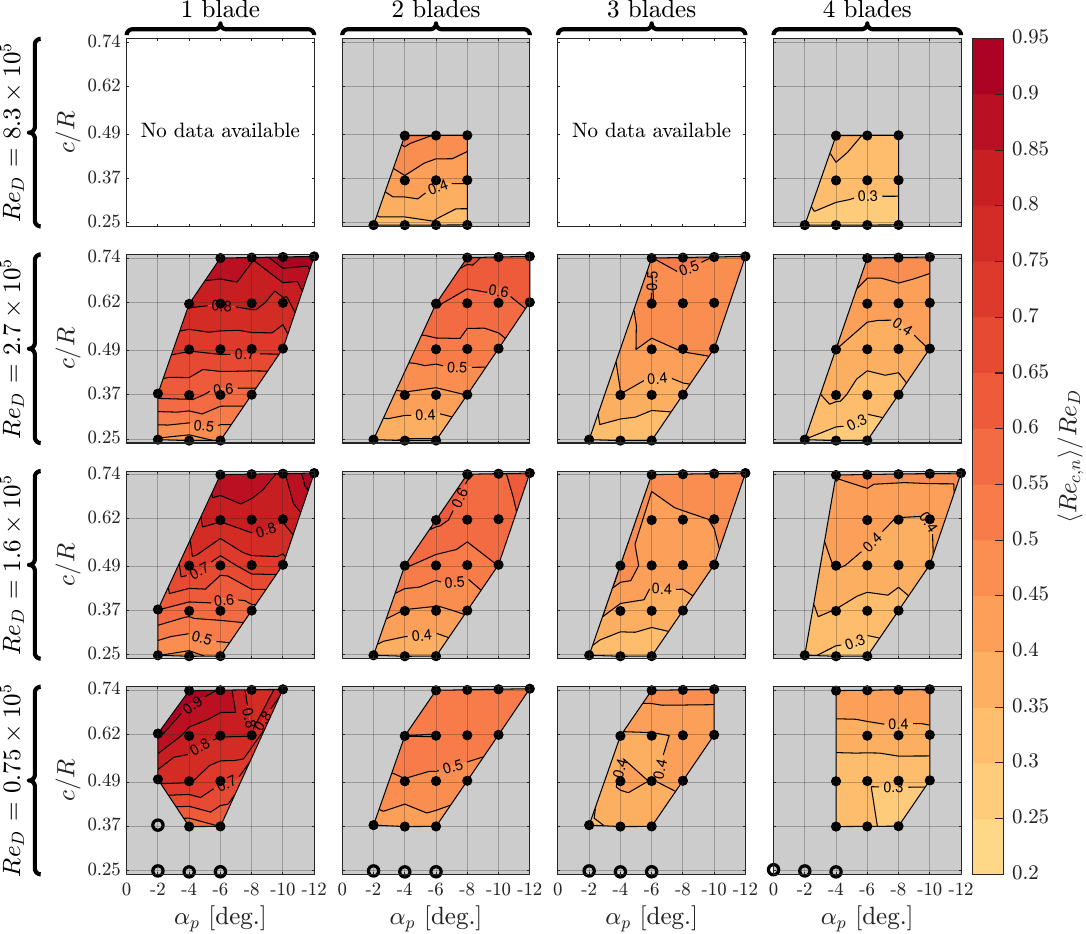}
    \caption{Contours of the cycle-average $Re_{c,n}$ at $\lambda_{\mathrm{opt}}$ normalized by the corresponding $Re_D$. Points with open circles indicate geometries that did not produce power at any tip-speed ratio and, as such, have no optimal operating condition.}
    \label{fig:reRatioHeatmap}
\end{figure*}

\noindent From this relation, it is observed that $Re_{c,n}$ increases with $Re_D$, $c/R$, and $\lambda$. However, these factors often move in competing directions. For example, as $c/R$ increases, $\lambda_{\mathrm{opt}}$ decreases (\Cref{fig:tsrHeatmap}), which decreases $Re_{c,n}$ at maximum efficiency.

To assess how turbine geometry influences $Re_{c,n}$ across the tested parameter space, the average value of \Cref{eq:reRelation} over a single rotation (i.e., $\langle Re_{c,n} \rangle$) is evaluated at $\lambda_{\mathrm{opt}}$ for each geometry. The resulting contours in \Cref{fig:reRatioHeatmap} suggest that, at $\lambda_{\mathrm{opt}}$, the largest $\langle Re_{c,n} \rangle$ for a given $Re_D$, $\alpha_{p}$, and $N$ is generally achieved by maximizing $c/R$.
In other words, at a given $Re_D$, the $c/R$ term in \Cref{eq:reRelation} has the stronger influence on $\langle Re_{c,n} \rangle$, with increasing $c/R$ more than offsetting the corresponding decrease in $\lambda_{\mathrm{opt}}$.
While this would be explanatory for why blades with the largest $c/R$ perform best at most $Re_D\!-\!N$ combinations (\Cref{fig:cpBladeHeatmap}), it is not explanatory for the inversion of this trend at high $Re_D$, where blades with lower $c/R$ perform best.
This contradictory result is indicative of the limitations of interpreting the observed performance trends via nominal $Re_c$. \Cref{eq:reRelation} neglects induction, which alters the relative velocity incident on the blade and thus the true local Reynolds number. Additionally, although it is known that increasing $c/R$ increases ``flow curvature'' experienced by the blade \citep{migliore_flow_1980,migliore_effects_1980,mandal_effects_1994}, how flow curvature alters $Re_c$ is not understood.
Interplay between these factors motivates further study of the influences of induction and flow curvature on blade-level fluid dynamics. 

\subsection{Effects of Geometric Parameters}
\label{disc:geom_params}

\subsubsection{Chord-to-Radius Ratio}
\label{disc:cToR}

Although the kinematics of cross-flow turbine blades are typically evaluated at a single point for convenience (e.g., the quarter-chord as in \Cref{fig:vec_dia}), the true kinematics will vary along the blade chord since the distance from the axis of rotation to each point along the blade profile varies. Consequently, the actual angle of attack and relative velocity vary with chord-wise position, causing a symmetric airfoil in curvilinear flow to perform like a cambered foil in rectilinear flow (``virtual camber") with an altered angle of the relative velocity (``virtual incidence") \citep{migliore_flow_1980, takamatsu_study_1985}.
The magnitude of these flow curvature effects increases with $c/R$, but also depends on the tip-speed ratio, preset pitch angle, and blade azimuthal position \citep{migliore_flow_1980}.
The presence of virtual camber and virtual incidence complicate design optimization since reduced order models for blade performance (e.g., blade-element momentum theory) typically utilize airfoil lift and drag coefficients as measured in rectilinear flow \citep{balduzzi_blade_2015}, although these models can include corrections for flow curvature \citep{bianchini_model_2011, bianchini_implementation_2018}.
Prior research has generally focused on conformal mapping techniques for determining the virtual camber and incidence for blades experiencing flow curvature \citep{migliore_flow_1980, balduzzi_blade_2015, rainbird_influence_2015, horst_flow_2016, bianchini_virtual_2016}. However, how flow curvature affects turbine performance metrics has received relatively limited attention outside of foundational work by \citet{migliore_flow_1980}, who tested two-bladed turbines with $c/R = 0.114$ and $c/R = 0.260$ at $Re_D \approx 2.5\times10^5 - 15\times10^5$ and attributed the lower performance of the $c/R = 0.260$ blades to flow curvature effects.

In contrast to the results of \citet{migliore_flow_1980}, for these experiments turbine geometries with the largest $c/R$---and thus the greatest flow curvature effects---yielded the highest $C_{P,\mathrm{blades}}$ of the tested geometries at many $Re_D\!-\!N$ combinations, particularly for lower $Re_D$ (\Cref{fig:cpBladeHeatmap}).
However, large $c/R$ blades were not universally optimal, and were outperformed by low $c/R$ blades at the highest $Re_D$ tested. As previously noted, this study did not determine the true optimal $c/R$ at each $Re_D\!-\!N$ combination and, in several cases the best-performing $c/R$ was either the minimum or maximum $c/R$ tested. Specifically, at the two lowest $Re_D$ tested, it appears that $c/R$ greater than 0.74 may benefit performance.
This is remarkable, given that, with the exception of the studies by \citet{shiono_experimental_2000} and \citet{el-samanoudy_effect_2010}, $c/R\!\approx\!0.74$ is larger than that tested in any previous experiments (\Cref{tab:prior_work}).
Similarly, at the highest tested $Re_D$, further reduction of $c/R$ below 0.25 may continue to improve performance, subject to structural limitations.
The improved performance of low $c/R$ blades at higher $Re_D$ agrees with the results of \citet{migliore_flow_1980} and is further supported by cross-comparison of previous studies in the Chase towing tank at $Re_D \approx 4\!\times\!10^{5}$ \ -- \ $13\!\times\!10^{5}$, which characterized three-bladed turbines (NACA 0020 airfoils) with $c/R = 0.10$ \citep{bachant_experimental_2016} and $c/R = 0.28$ \citep{bachant_characterising_2015, bachant_effects_2016}. 
\citet{bianchini_design_2015} also identified a similar change in the optimal $c/R$ with increasing freestream velocity using a numerical method based on Blade Element Momentum theory, although the turbine diameter (and thus $Re_D$) was allowed to vary while the projected area ($HD$) was held constant.
It is recommended that future work broaden the range of $c/R$ and $Re_D$ explored to identify the limits of these trends.

\subsubsection{Blade Count}
\label{disc:bladeCount}

\begin{figure*}[t]
    \centering
    \includegraphics[width=\textwidth]{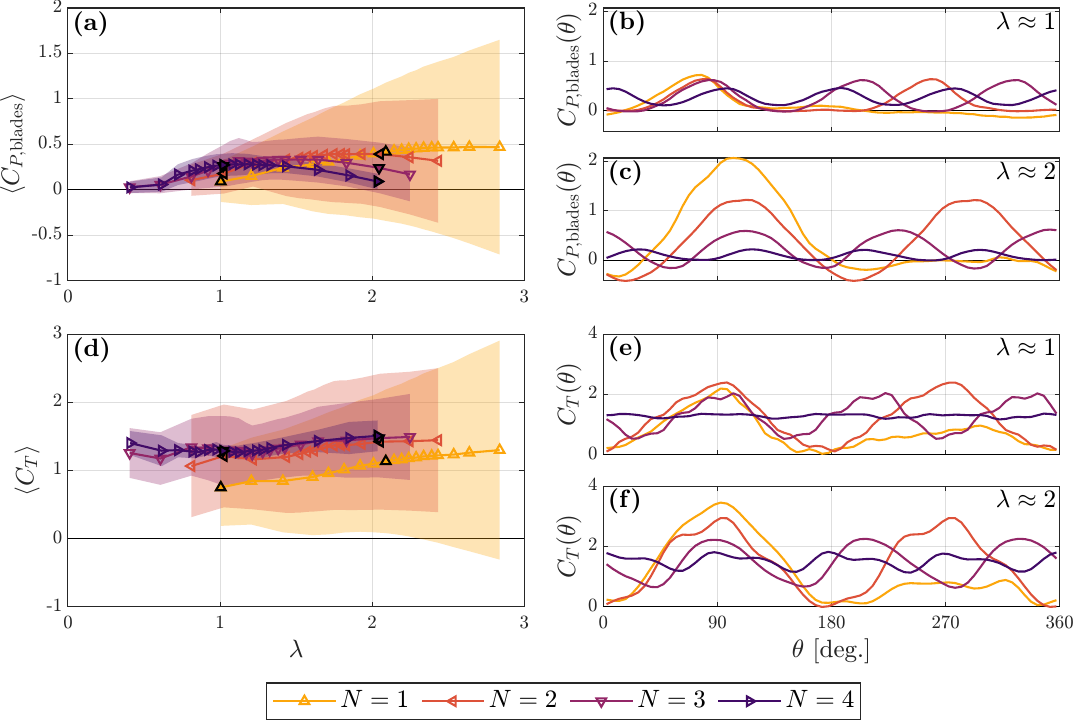}
    {\phantomsubcaption\label{fig:bladeTimeCP}
    \phantomsubcaption\label{fig:bladePhaseCP1}
    \phantomsubcaption\label{fig:bladePhaseCP2}
    \phantomsubcaption\label{fig:bladeTimeCT}
    \phantomsubcaption\label{fig:bladePhaseCT1}
    \phantomsubcaption\label{fig:bladePhaseCT2}}
    
    \caption{Influence of blade count on performance for $c/R = 0.62$, $\alpha_p = -6^{\circ}$, $Re_D = 2.7\times10^5$. \subref{fig:bladeTimeCP} Time-average $C_{P,\mathrm{blades}}$ as a function of $\lambda$. The shaded region indicates $\pm1$ standard deviation of the values measured during the median rotational cycle (i.e., $C_{P,\mathrm{blades}}(\theta)$) at each $\lambda$ set-point. Phase-median $C_{P,{\mathrm{blades}}}$ is shown for \subref{fig:bladePhaseCP1} $\lambda \approx 1$ and \subref{fig:bladePhaseCP2} $\lambda \approx 2$. \subref{fig:bladeTimeCT} Time-average $C_T$ as a function of $\lambda$ with the shaded region indicating $\pm1$ standard deviation as for \subref{fig:bladeTimeCP}. Phase-median $C_{T}$ is shown for \subref{fig:bladePhaseCT1} $\lambda \approx 1$ and \subref{fig:bladePhaseCT2} $\lambda \approx 2$. }
    \label{fig:bladeTrends}
\end{figure*}

Increasing the number of turbine blades for fixed $\alpha_p$ and $c/R$ decreases the maximum blade-level efficiency (\Cref{fig:cpBladeHeatmap}) and the tip-speed ratio at which maximum efficiency occurs (\Cref{fig:tsrHeatmap}).
While decreases in maximum $C_{P,\mathrm{blades}}$ and $\lambda_{\mathrm{opt}}$ with increasing $N$ are well-documented in prior work \citep{mcadam_experimental_2013, li_effect_2015, araya_transition_2017, miller_solidity_2021}, it is observed that increasing $N$ from 1 to 2 actually increases maximum $C_{P,\mathrm{blades}}$ for a few cases at low $Re_D$ and low $c/R$ (e.g., $Re_D = 0.75 \times 10^{5}$ and $c/R = 0.37$).
Additionally, away from $\lambda_{\mathrm{opt}}$, the relationship between $N$ and $C_{P,\mathrm{blades}}$ varies with $\lambda$: as shown for a subset of geometries ($c/R = 0.62$, $\alpha_p = -6^{\circ}$, and $Re_D = 2.7\times10^5$) in \Cref{fig:bladeTimeCP}, the most efficient blade count increases as $\lambda$ decreases.
At lower $\lambda$, (e.g., $\lambda \approx 1$ in \Cref{fig:bladeTimeCP}), turbines with more blades have higher time-average efficiencies, whereas at higher $\lambda$ (e.g., $\lambda \approx 2$ in \Cref{fig:bladeTimeCP}), turbines with fewer blades become most efficient.
In contrast, the time-average thrust on the rotor generally increases with $N$ for all $\lambda$ (\Cref{fig:bladeTimeCT}).
Based on these results, it is hypothesized that $N$ primarily influences performance by altering near-blade induction. 
At a given $\lambda$ and $c/R$, as $N$ increases, the rotor appears more solid to the freestream, and thus diverts more flow around the turbine \citep{araya_transition_2017}.
Consequently, the velocity incident on the blades at a given $\lambda$ is expected to decrease, which would reduce the power that can be produced by an individual blade at that $\lambda$.
This hypothesis is supported by the results in \Cref{fig:bladePhaseCP1,fig:bladePhaseCP2}, where the maxima for individual blade power (periodic peaks) decrease as $N$ increases.
Since induction increases with rotation rate \citep{araya_transition_2017}, the severity of this power-per-blade penalty varies with $\lambda$.
For example, at $\lambda \approx 1$ (\Cref{fig:bladePhaseCP1}), the power maxima are only marginally reduced as $N$ increases, whereas at $\lambda \approx 2$ (\Cref{fig:bladePhaseCP1}), the power maxima steeply decline with each blade added.
It is noted that, while induction is also expected to increase with $c/R$ for similar reasons as increasing $N$, the chord-to-radius ratio influences performance through additional mechanisms (e.g., flow curvature, $Re_c$ length scale) that are absent for $N$.
Additionally, since cross-flow turbines experience both axial and lateral induction, caution is warranted in attempting to correlate induction directly to thrust force, as for axial flow turbines \citep{burton_wind_2011}.
Future research should quantitatively assess the relationship between turbine geometry, performance, and induction via flow visualization.

Over the course of a rotational cycle, variation in $C_{P,\mathrm{blades}}$ (e.g., \Cref{fig:bladePhaseCP1,fig:bladePhaseCP2}) and $C_T$ (e.g., \Cref{fig:ctVarHeatmap,fig:bladePhaseCT1,fig:bladePhaseCT2}) is significantly reduced as $N$ increases. As indicated by the shaded regions of \Cref{fig:bladeTimeCP,fig:bladeTimeCT}, this effect is especially pronounced as $\lambda$ increases.
Consequently, the choice of blade count constitutes a trade off between 1) maximum efficiency and 2) cyclic variation of power generation and structural loads.
A turbine with fewer blades can produce power over a wider range of tip-speed ratios and achieve higher maximum efficiency, but will also be subject to more intense cyclic loading and more susceptible to fatigue.
Consequently, while single-bladed turbines are useful for highlighting trends across this parameter space and isolating blade-level dynamics, they are likely impractical in most applications.
Conversely, for $N=4$, the amplitude of cyclic variation is significantly reduced, but at the cost of cycle-average efficiency. Even higher blade counts would likely only offer incremental gains in consistency at the cost of further reductions in efficiency. Therefore, two- and three-bladed turbines may represent a feasible compromise between cycle-average power and cyclic variation.

\subsubsection{Preset Pitch Angle}
\label{disc:presetPitch}

As $c/R$ increases, the $\alpha_{p}$ that yields maximum time-average $C_{P,\mathrm{blades}}$ generally becomes more negative (i.e., more toe-out) (\Cref{fig:pitch_CToR_Lineplot}). The preset pitch angle influences blade fluid dynamics by changing the angle of attack at each azimuthal position, and thus, the blade's lift and drag coefficients. Considering the nominal angle of attack trajectories (\Cref{fig:nomAlpha}), a more negative $\alpha_{p}$ is expected to decrease the maximum nominal angle of attack the blade encounters during the upstream sweep ($0^{\circ} \leq \theta < 180^{\circ}$), delaying the onset of dynamic stall and reducing its severity. Therefore, previous studies have hypothesized that appropriately decreasing the preset pitch angle increases the net power produced during the upstream sweep by mitigating the abrupt loss of lift and increase in drag caused by boundary layer separation during dynamic stall \citep{strom_consequences_2015, du_experimental_2019}.

\begin{figure*}[t]
    \centering
    \includegraphics[width=\textwidth]{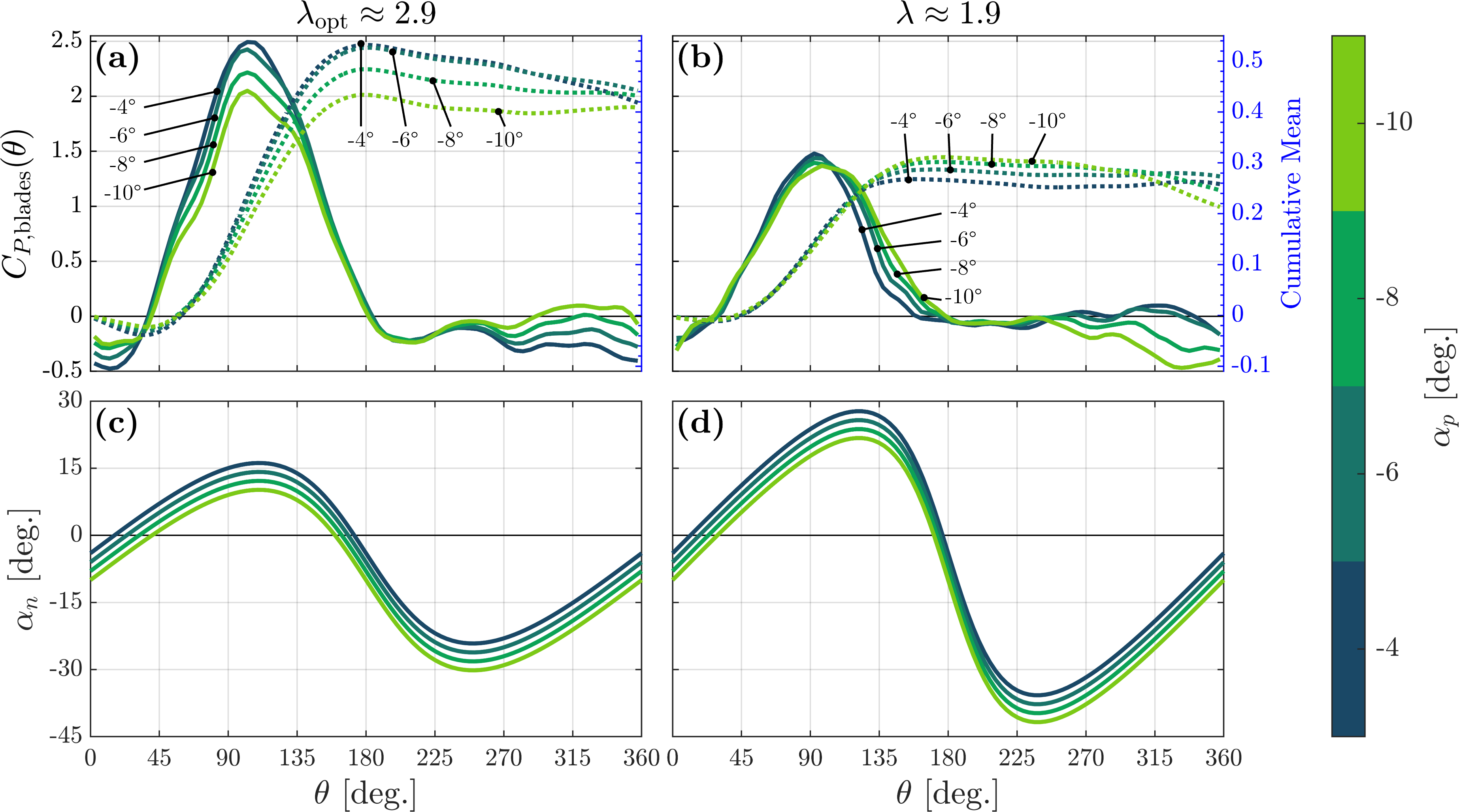}

    {\phantomsubcaption\label{fig:pitchPhaseOpt}
    \phantomsubcaption\label{fig:pitchPhaseLow}
    \phantomsubcaption\label{fig:pitchAlphaOpt}
    \phantomsubcaption\label{fig:pitchAlphaLow}}
    
    \caption{\subref{fig:pitchPhaseOpt} Phase-median $C_{P,\mathrm{blades}}$ as a function of azimuthal position for various $\alpha_{p}$ at $c/R = 0.49$, $N=1$, $Re_D = 2.7\times10^5$, and $\lambda_{\mathrm{opt}} \approx 2.9$. Dashed lines show a cumulative average of $C_{P,\mathrm{blades}}$ over the cycle (i.e., average $C_{P,\mathrm{blades}}$ between 0$^\circ$ and $\theta$). \subref{fig:pitchPhaseLow} Phase-median and cumulative average $C_{P,\mathrm{blades}}$ for the same turbines operating at $\lambda \approx 1.9$. The preset pitch angle for each curve is annotated and indicated by color. The nominal angle-of-attack trajectories for various $\alpha_p$ are shown for \subref{fig:pitchAlphaOpt} $\lambda \approx 2.9$ and \subref{fig:pitchAlphaLow} $\lambda \approx 1.9$.}
    \label{fig:pitchPhase}
\end{figure*}

To explore the fluid dynamics that drive trends in $\alpha_p$, a representative set of single-bladed turbines---for which the individual blade power can be isolated within a cycle---are considered at $c/R = 0.49$ and $Re_D = 2.7\times10^5$.
Phase-median $C_{P,\mathrm{blades}}$ and corresponding nominal angle of attack trajectories for these turbines are visualized at two tip-speed ratios: the optimal tip-speed ratio, at which light dynamic stall is expected ($\lambda_{\mathrm{opt}} \approx 2.9$; \Cref{fig:pitchPhaseOpt,fig:pitchAlphaOpt}), and a lower tip-speed ratio at which deeper dynamic stall is expected ($\lambda \approx 1.9$; \Cref{fig:pitchPhaseLow,fig:pitchAlphaLow}).
At both $\lambda$ shown, the highest instantaneous $C_{P,\mathrm{blades}}$ is achieved with the least negative (i.e., least toe-out) tested preset pitch of $\alpha_p = -4^{\circ}$, which corresponds to the highest maximum $\alpha_n$ during the upstream sweep (\Cref{fig:pitchAlphaOpt,fig:pitchAlphaLow}).
For the $\lambda_{\mathrm{opt}} \approx 2.9$ case, $\alpha_p = -4^{\circ}$ also yields the greatest average power over the upstream sweep (\Cref{fig:pitchPhaseOpt}). While this is contrary to the hypothesized benefits of delaying dynamic stall, the importance of dynamic stall is more limited at this tip-speed ratio \citep{snortland_cycletocycle_2023}. 
In contrast, for the $\lambda \approx 1.9$ case, a more negative (i.e., more toe-out) pitch angle of $\alpha_p = -10^{\circ}$ yields the greatest average power over the upstream sweep (\Cref{fig:pitchPhaseLow}). Although $\alpha_p = -10^{\circ}$ results in the lowest peak instantaneous power at this $\lambda$, power is produced over a wider range of $\theta$ in the upstream sweep, consistent with the hypothesis of a more negative pitch delaying dynamic stall via reduction of the maximum $\alpha_n$ (\Cref{fig:pitchPhaseLow}).

Since blades with larger $c/R$ operate at lower tip-speed ratios (\Cref{fig:tsrHeatmap}) and thus experience deeper dynamic stall, these blades may benefit from more negative preset pitch relative to blades with smaller $c/R$.
However, for both $\lambda$, neither $\alpha_p = -4^{\circ}$ nor $\alpha_p = -10^{\circ}$ yields the maximum cycle-average power. Instead, the cumulative phase-median performance profiles (\Cref{fig:pitchPhaseOpt,fig:pitchPhaseLow}) show that the optimal $\alpha_p$ (in this case, $-6^{\circ}$) maximizes cycle-average power by balancing power production during the upstream sweep with power consumption during the downstream sweep ($180^{\circ} \leq \theta < 360^{\circ}$) as the blade moves through the wake. Consequently, an explanation of $\alpha_p$ trends that focuses only on the upstream sweep is incomplete.
Unfortunately, trends in the downstream sweep cannot be explained by the nominal angle of attack trajectories (\Cref{fig:pitchAlphaOpt,fig:pitchAlphaLow}), since induction is particularly significant during the downstream sweep.
Additionally, both induction and flow curvature vary with $\lambda$ and $\theta$ and cause the actual angle of attack to deviate from the nominal value, complicating a holistic interpretation of the effects of the preset pitch angle. It is recommended that future research focus on identifying the fluid dynamic basis for these performance trends.

\subsubsection{Solidity}
\label{disc:solidity}

\begin{figure*}[t]
    \centering
    \includegraphics[width=\textwidth]{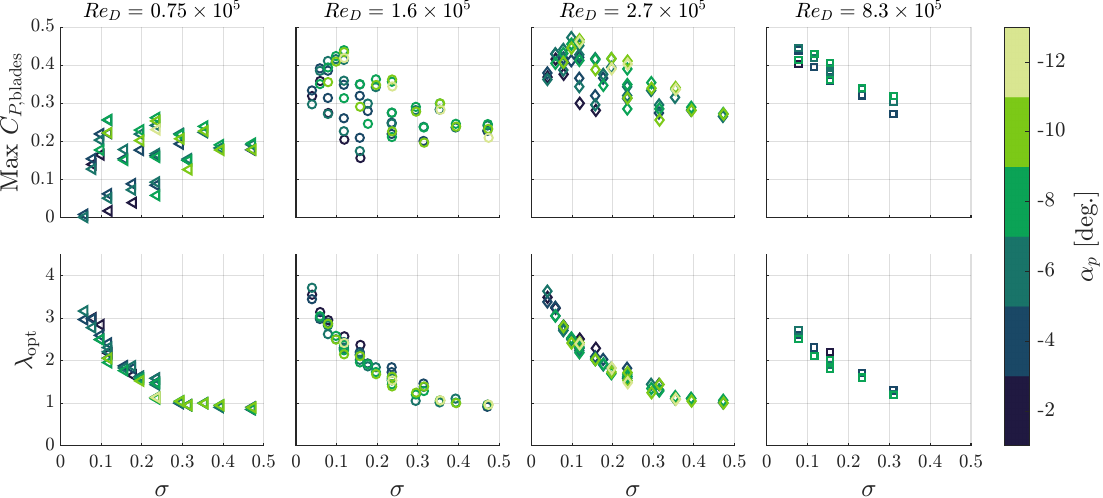}
    \caption{Maximum $C_{P,\mathrm{blades}}$ and associated $\lambda_{\mathrm{opt}}$ and as a function of solidity for each $\alpha_p$ and $Re_D$.}
    \label{fig:solidityPerf}
\end{figure*}

As discussed in \Cref{sec:introduction}, it is relatively common to combine $c/R$ and $N$ to form a single dimensionless parameter, solidity, $\sigma$ (\Cref{eq:solidity}), that represents the fraction of the rotor circumference containing blades.
For a given $\lambda$, as $\sigma$ increases, less flow is able to pass through the turbine since the blades present greater resistance to the flow. At the extreme case of $\sigma = 1$, the entire rotor circumference would be occupied by the blades and the turbine would approximate a cylindrical shell. The $c/R$ -- $N$ combinations in this work resulted in an order of magnitude range of solidity ($\sigma \approx 0.04\!$ -- $\! 0.47$). Additionally, by simultaneously varying these parameters, the same value of $\sigma$ was achieved with different combinations of $c/R$ and $N$ (e.g., doubling $N$ and halving $c/R$). This allows the suitability of solidity as a descriptive dimensionless parameter to be assessed.

As shown in \Cref{fig:solidityPerf}, holding solidity constant does not guarantee constant characteristic performance. At low-to-moderate $\sigma$, turbine geometries with similar solidity exhibit a wide range of maximum  $C_{P,\mathrm{blades}}$, even when both $Re_D$ and $\alpha_p$ are held constant. 
This is not unexpected, given that $c/R$ and $N$ affect $C_{P,\mathrm{blades}}$ through distinct mechanisms. For example, if $\sigma$ is increased by increasing $N$ only, the local Reynolds number can change due to variations in induction. However, if $\sigma$ is increased by increasing $c/R$ only, then in addition to changes in the local Reynolds number due to changes in induction and length scale ($c$), flow curvature also increases. Consequently, for two turbines with the same $\sigma$, but different $c/R$ and $N$, the underlying fluid dynamics for the two turbines are unlikely to be equivalent, nor are the associated fluid dynamics likely to scale in the same way with Reynolds number. While not shown, a similarly wide range is observed for $C_T$ and $C_L$ at $\lambda_{\mathrm{opt}}$ for a given $\sigma$.

Due to the unique hydrodynamic influences of $c/R$ and $N$, cross-comparisons made solely on the basis of solidity can be misleading, and likely contribute to contradictory trends across both experimental and numerical prior studies regarding the effect of solidity on maximum $C_P$.
Multiple studies report that maximum $C_{P}$ decreases as $\sigma$ increases \citep{howell_wind_2010, mcadam_experimental_2013, li_effect_2016, lohry_unsteady_2016, araya_transition_2017, miller_solidity_2021}, which is expected for an increase in $N$ (\Cref{fig:cpBladeHeatmap}).
In contrast, several other studies find that maximum $C_{P}$ increases with increasing solidity \citep{consul_influence_2009, goude_simulations_2014, eboibi_experimental_2016, rezaeiha_optimal_2018, lam_assessment_2018, blackwell_selected_1977}, an effect that is observed in this work for increasing $c/R$ at lower $Re$ (\Cref{fig:cpBladeHeatmap}).
Across prior studies, $\sigma$ is not consistently varied: some studies changed $\sigma$ only by varying $N$ \citep{consul_influence_2009, howell_wind_2010, mcadam_experimental_2013, li_effect_2016, araya_transition_2017, lam_assessment_2018, miller_solidity_2021}, others varied $c/R$ \citep{shiono_experimental_2000, goude_simulations_2014, bianchini_design_2015, lohry_unsteady_2016, eboibi_experimental_2016, rezaeiha_optimal_2018, du_experimental_2019}, and still others varied both $N$ and $c/R$ \citep{blackwell_selected_1977, li_numerical_2010, priegue_influence_2017}.
While a few studies hold solidity constant through simultaneous variation of $c/R$ and $N$ \citep{blackwell_selected_1977, shiono_experimental_2000}, no prior studies have systematically demonstrated that identical performance is obtained for a given $\sigma$, regardless of the values of $c/R$ and $N$.
However, when $\sigma$ is used as a dimensionless parameter for performance characterization, this assumption is seldom questioned. For example, with the exception of the study by \citet{miller_solidity_2021} there appears to be no acknowledgment that changing $\sigma$ via $c/R$ rather than $N$ could introduce flow curvature effects, even though these effects have been documented since 1980 \citep{migliore_flow_1980}. 

This is not, however, to say that solidity is a universally poor parameter for describing cross-flow turbine performance. At the highest solidities tested (corresponding to geometries with both high $c/R$ and high $N$), turbines with similar $\sigma$ have similar maximum $C_{P,\mathrm{blades}}$ (i.e., performance is relatively insensitive to $\alpha_{p}$). Similar results are observed at $Re_D = 8.3\times10^5$, which is consistent with a hypothesis that at high Reynolds numbers and low $c/R$, the flow curvature and Reynolds number effects linked to $c/R$ are negligible, and solidity is thus a good predictor of $C_P$. This hypothesis is supported by foundational research by  \citet{blackwell_selected_1977}, who held $\sigma$ constant by varying $c/R$ and $N$ simultaneously and observed similar---but not identical---characteristic performance at high $Re_D$ ($\approx 3.6 - 41 \times 10^{5}$) and low $c/R$ ($< 0.1$). They were unable to identify the root cause for the small disagreement in performance, which is unsurprising, given that these studies took place several years prior to the first documentation of flow curvature effects. 

\begin{figure}[t]
    \centering
    \includegraphics[width=0.4\textwidth]{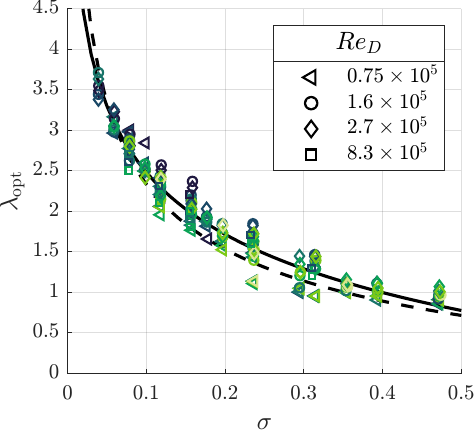}
    \caption{$\lambda_{\mathrm{opt}}$ versus solidity for all turbines tested. Marker color indicates $\alpha_p$ as in \Cref{fig:solidityPerf}. The curve fit in \Cref{eq:solidityFit} is shown as a solid black line, whereas the curve fit from \citet{rezaeiha_optimal_2018} is shown as the dashed black line.}
    \label{fig:solidityRegression}
\end{figure}

Notably, $\lambda_{\mathrm{opt}}$ varies closely with solidity since increases in $c/R$ and $N$ both decrease $\lambda_{\mathrm{opt}}$, regardless of $\alpha_p$ or $Re_D$ (\Cref{fig:solidityPerf}).
This inverse relationship between $\sigma$ and $\lambda_{\mathrm{opt}}$ has been well documented in prior experimental studies \citep{shiono_experimental_2000, howell_wind_2010, mcadam_experimental_2013, li_effect_2015, eboibi_experimental_2016, araya_transition_2017, miller_solidity_2021} and simulation studies \citep{consul_influence_2009, li_numerical_2010, goude_simulations_2014, lohry_unsteady_2016, rezaeiha_optimal_2018}, but this work expands the geometric and operational space over which this is observed. Furthermore, the relationship between $\lambda_{\mathrm{opt}}$ and $\sigma$ suggests that turbines with the same ``static'' solidity ($\sigma$) have similar optimal dynamic solidities as defined by \citet{araya_transition_2017}.
Overlaying $\lambda_{\mathrm{opt}}$ versus $\sigma$ for all tested geometries as in \Cref{fig:solidityRegression} illustrates that, across the parameter space tested, $\lambda_\mathrm{opt}$ is determined primarily by $\sigma$, with $\alpha_p$ and $Re_D$ having only secondary effects, particularly at the extrema of the tested $Re_D$.
This is consistent with the relative insensitivity of $\lambda_{\mathrm{opt}}$ to $Re_D$ and $\alpha_p$ observed in \Cref{fig:tsrHeatmap}, as well as the $Re_D$-independence of $\lambda_{\mathrm{opt}}$ observed in prior work \citep{rezaeiha_characterization_2018, bachant_effects_2016, miller_verticalaxis_2018}.
Additionally, some spread in $\lambda_{\mathrm{opt}}$ at a given $\sigma$ is likely attributable to differences in how finely $\lambda_{\mathrm{opt}}$ was resolved across tests.
Following \citet{rezaeiha_optimal_2018}, who observed a similar trend between $\sigma$ and $\lambda_{\mathrm{opt}}$ for $\sigma \approx 0.03 - 0.36$, the data in \Cref{fig:solidityRegression} are fit to a two-term power law of the form
%
\begin{equation} 
    \lambda_{\mathrm{opt}} = 9.882\sigma^{-0.093} - 9.764 .
    \label{eq:solidityFit}
\end{equation}
\noindent The result (solid line in \Cref{fig:solidityRegression}) is similar to that obtained by \citeauthor{rezaeiha_optimal_2018} (dashed line), although \citeauthor{rezaeiha_characterization_2018}'s equation yields slightly lower values of $\lambda_{\mathrm{opt}}$ for the range of solidities tested here.

Considering these results as a whole, while solidity may capture trends in $\lambda_{\mathrm{opt}}$ and, in some cases, $C_{P,\mathrm{blades}}$, it is clear from \Cref{fig:solidityPerf} that solidity is an incomplete descriptor of turbine dynamics. Consequently, the use of solidity as a representative dimensionless quantity is discouraged and future research is recommended to differentiate the overlapping and unique influences of $c/R$ and $N$ on performance.

\subsection{Design Considerations}
\label{disc:design}
While the trends in $Re_D$, $c/R$, $N$, and $\alpha_p$ are discussed from a fluid dynamics perspective in \Cref{disc:geom_params}, in a field-scale application the ultimate choice of turbine geometry will be informed by design objectives. For example, while operating at the highest $Re_D$ possible will maximize $C_{P,\mathrm{blades}}$ for any geometry, $Re_D$ is typically set by site conditions (e.g., expected inflow velocities) as well as physical and economic constraints on rotor dimensions. Similarly, in addition to material costs, structural limitations will dictate the choice of $c/R$ (e.g., minimum allowable chord length at high $Re_D$) and number of blades (e.g., maximum allowable amplitude of cyclic loading). In contrast, from a design perspective, $\alpha_p$ is a free parameter and may be chosen simply to maximize efficiency based on the selected $Re_D$, $c/R$ and $N$.

$\lambda_{\mathrm{opt}}$ for each turbine provides a convenient point of reference for visualizing trends across the broad parameter space spanned by these experiments (e.g., as in \Cref{fig:cpBladeHeatmap}). This does, however, obscure a considerable amount of information about each turbine's performance over the full range of operating conditions (i.e., characteristic performance away from the optimal point, such as that briefly explored in \Cref{fig:bladeTrends}), as well as the variations that occur within a cycle at a given $\lambda$ set point (e.g., \Cref{fig:phaseAveSynth,fig:bladeTrends,fig:pitchPhase}). This information plays an important role in co-design. 
First, for economic reasons driven by capacity utilization, wind and water turbines do not operate at $\lambda_{\mathrm{opt}}$ for all inflow conditions. Instead, beyond a ``rated'' inflow condition, generation systems often enter a mode of approximately constant power \citep{burton_wind_2011}. For fixed-pitch turbines, this can be achieved via the intentional reduction of $C_P$ by operating at a higher- or lower-than-optimal $\lambda$. As such, characteristic performance at other operating points is relevant to design.
Second, turbine structural and driveline components (e.g., generator) are sized based on the maximum, not cycle-average, expected loads. The magnitudes of these cyclic loads will affect cost and may affect component efficiency within a cycle, as well as fatigue life.
To support design efforts, as well as numerical model validation, the full data archive---including both time-average and phase-median performance for each geometry---is openly available through Dryad \citep{hunt_data_2024}. 

\subsection{Limitations and Knowledge Gaps}
\label{sec:gaps}
While the 223 experiments conducted in this study provide insights into the interplay between $c/R$, $\alpha_{p}$, $N$, and $Re_D$, some trends are not fully resolved.
Foremost, as described in \Cref{disc:cToR}, despite the wide range of $c/R$ tested (0.25-0.74), the true optimal $c/R$ was not resolved for every $Re_D\!-\!N$ combination. 
Furthermore, interpretation of $c/R$ trends is also complicated by the influence of flow curvature effects (virtual camber and virtual incidence), which would benefit from further experimental investigation.
Additionally, while the optimal preset pitch angle was captured at each $c/R$ for all $Re_D\!-\!N$ combinations, the $\alpha_{p}$ that were tested at each $c/R$ were limited for expediency to those around the optimal value. Additional tests farther from the optimal value of preset pitch could illuminate how $\alpha_{p}$ influences power production and flow recovery across $c/R$ and $\lambda$.

\begin{figure}[t]
    \centering
    \includegraphics[width=0.8\textwidth]{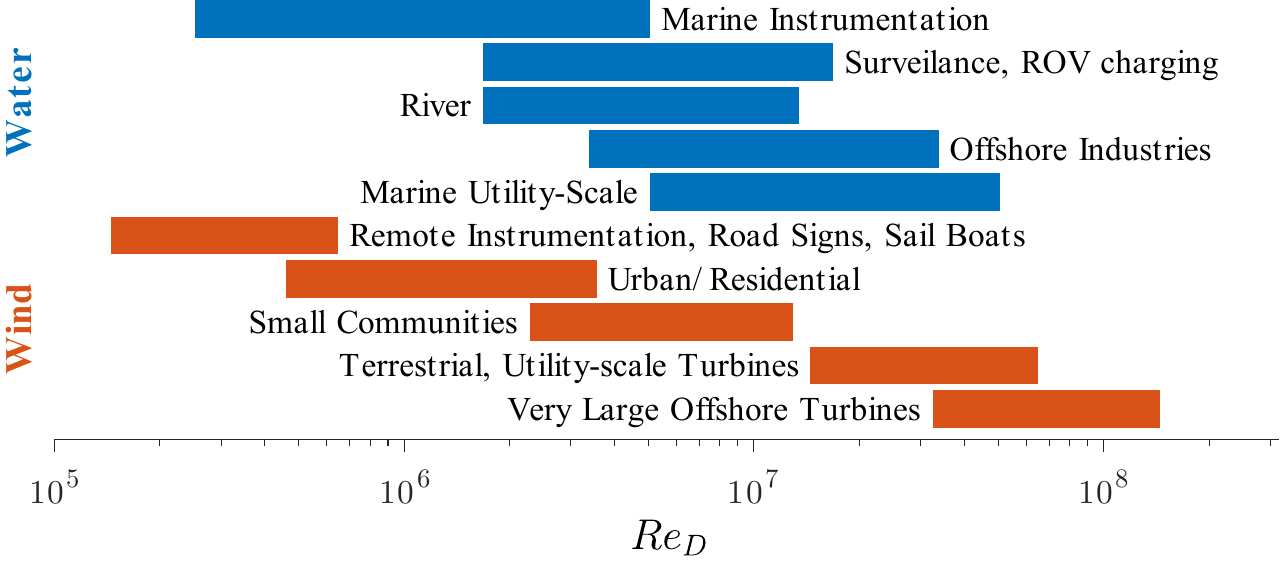}
    \caption{Ranges of expected diameter-based Reynolds numbers for cross-flow turbines by application.}
    \label{fig:re_scales}
\end{figure}

Although the $Re_D$ considered in this study span an order of magnitude, experiments at UW were conducted at transitional Reynolds numbers ($Re_D = 0.75\times10^5 - 2.7\times10^5$), and it is unknown whether Reynolds independence was achieved at UNH ($Re_D = 8.3\times10^5$).
Consequently, trends in performance with geometry may differ outside of the range of $Re_D$ tested in this study, although certain trends appear insensitive to $Re_D$ (e.g., \Cref{fig:ctVarHeatmap,fig:solidityRegression}).
Since, in general, the absolute power output from a cross-flow turbine scales with physical dimension, it is natural to question the value of exploring these parameters for cross-flow turbines at relatively low Reynolds numbers.
However, as shown in \Cref{fig:re_scales}, the range of relevant values of $Re_D$ for the full application space in wind and water spans several orders of magnitude, and distributed power generation across these scales is critical for meeting international energy and climate goals \citep{livecchi_powering_2019, united_nations_general_assembly_transforming_2015}.
The $Re_D$ tested in this study overlap with that of smaller vertical-axis wind turbines in the field \citep{dabiri_potential_2011} that still exhibit Reynolds-dependence \citep{ miller_verticalaxis_2018}, and cross-flow turbines can still perform well at small physical scales (i.e., low $Re_D$ in \Cref{fig:cpBladeHeatmap}).
Further, from an experimental perspective, unless compressed-air wind tunnels are employed \citep{miller_verticalaxis_2018, miller_solidity_2021}, turbines that are small enough to test in laboratories will necessarily have low Reynolds numbers. 
Similarly, direct numerical simulations of cross-flow turbines, which can provide additional insight into blade-level fluid dynamics, are often limited to relatively low Reynolds numbers by computational cost.
Here, experimental data at low Reynolds numbers can be helpful for validation. 

Finally, several geometric parameters that impact performance were not considered here. These include the blade profile (e.g., other symmetric foils, cambered foils), blade cant angle, spanwise profile (e.g., helical blades), and surface roughness. While the trends observed in this work may change if, for example, all experiments were repeated with cambered blades, the portion of the parameter space explored here lays a foundation for interpreting the results of future studies.

\section{Conclusions}
\label{sec:conclusion}
The geometric configuration of a cross-flow turbine influences blade fluid dynamics, and thus the power produced and loading experienced. Here, the interplay between three geometric parameters---the chord-to-radius ratio, the preset pitch angle, and number of blades---is considered across diameter-based Reynolds numbers spanning an order of magnitude ($\approx 8\times10^4 - 8\times10^5$). Expanding upon prior research, 223 unique experiments were conducted, from which the following key trends emerge:

\begin{enumerate}
    \item The Reynolds number has the greatest overall effect on turbine efficiency across all geometries. Until Reynolds independence is reached, increasing  $Re_D$ will increase $C_{P,\mathrm{blades}}$ for a given geometry.
    \item The optimal $c/R$ is primarily determined by the Reynolds number. At low $Re_D$, blades with large $c/R$ generate the most power. As $Re_D$ increases, blades with low $c/R$ are more efficient. The number of blades has a secondary effect on optimal $c/R$, but this effect also depends on the Reynolds number.
    \item Across all geometries, moderate toe-out preset pitch substantially improves efficiency (e.g., $-4^{\circ}$ to $-10^{\circ}$ in this study). In general, as a turbine's chord-to-radius ratio increases, the $\alpha_p$ that yields maximum efficiency becomes more negative.
    \item As $N$ increases, maximum efficiency decreases, and net loading of the turbine rotor increases; however the amplitude of cyclic loading decreases with each additional blade.  As such, the choice of $N$ constitutes a trade-off between maximum power and consistency of power and loading within a cycle.
    \item Solidity (\Cref{eq:solidity}) does not capture trends in $C_{P,\mathrm{blades}}$ for the range of $c/R$ and $Re_D$ investigated. Turbines with the same solidity, but different $c/R$ and $N$, can have different efficiencies. However, $\lambda_{\mathrm{opt}}$ is strongly correlated with solidity---turbines with greater chord-to-radius ratios and/or more blades tend to operate at slower tip-speeds. Therefore, outside of $\lambda_{\mathrm{opt}}$, the use of solidity as a non-dimensional parameter for characterizing turbine performance is discouraged.
\end{enumerate}

 Importantly, these results provide a uniquely holistic view of how the effects of these geometric parameters combine to influence turbine performance across a broad parameter space. 
 Given these performance trends and the inter-dependencies observed, the following design strategies are suggested.
 First, for small turbine sizes or low freestream velocities, where the chord-based Reynolds number is in the transitional regime, an optimal turbine geometry will consist of few blades with relatively large chord-to-radius ratios. This type of rotor, which has a high solidity, will operate at lower tip-speed ratios to maintain optimal induction. As dynamic stall is more significant at lower tip-speeds, these turbines perform best with larger toe-out preset pitch angles (e.g., $-6^{\circ}$ to $-10^{\circ}$ in this study).
 Second, optimal turbine geometries at large scales, well above the transitional chord-based Reynolds number, will benefit from a higher blade count with a smaller chord-to-radius ratio. Since these rotors have lower solidity, they will tend to operate at higher tip-speed ratios, where dynamic stall is less severe and smaller toe-out preset pitch angles provide the greatest cycle-average efficiency.

This work focuses primarily on trends at maximum-performance across the parameter space, but turbine design is informed by performance at off-optimal conditions, as well. Additionally, certain mechanisms, such as the specific effects that flow curvature has on turbine performance or how the preset pitch angle influences near-blade hydrodynamics during the blade's downstream sweep, require further investigation. Nonetheless, this study provides a rich data set for interpreting turbine fluid dynamics, validating reduced-order models and numerical simulations, informing turbine design, and unifying apparently contradictory trends in prior studies. 

\section{Acknowledgements}
Funding for this research was provided by the U.S. Navy's Naval Facilities Engineering and Systems Command (NAVFAC) under N0002410D6318 / N0002418F8702. Support from the Alice C. Tyler Charitable Trust is gratefully acknowledged for upgrades to the flume at UW that allow higher flow rates, heating, and cooling. Hannah Ross was supported by a National Science Foundation Graduate Research Fellowship (DGE-1256082). 

The authors would like to thank Peter Bachant for developing the turbine test bed at the University of New Hampshire, as well as guidance during data acquisition.
Abigale Snortland, Ari Athair, Owen Williams, and Robert Thresher are thanked for several insightful discussions related to the effects of the geometric parameters studied in this work.
Many thanks to Mark Miller, Thorsten Stoesser, Robert Howell, and Daniel Araya for their help in populating Table~\ref{tab:prior_work} with parameters omitted in the underlying publications.

\section{Data Availability}
The dataset supporting this work, as well as additional visualizations of the parameter space, are openly available in Dryad's digital repository, at \url{http://hdl.handle.net/1773/50607} \citep{hunt_data_2024}.

\printcredits


\appendix
\numberwithin{equation}{section}
\numberwithin{figure}{section}
\numberwithin{table}{section}

\section{Experimental Uncertainty}
\label{appendix:repeatability}

\begin{table*}[t]
    \caption{Specifications and systematic standard uncertainty for the instruments used to measure turbine efficiency.}
    \resizebox{\textwidth}{!}{ 
            \begin{threeparttable}
            \begin{tabular}{@{}ccccl@{}}
            \toprule
            \textbf{Facility} & \textbf{Instrument} & \textbf{\begin{tabular}[c]{@{}c@{}}\boldmath{$C_P$}\\ component\end{tabular}} & \textbf{\begin{tabular}[c]{@{}c@{}}Measured\\ Quantity\end{tabular}} & \multicolumn{1}{c}{\textbf{\begin{tabular}[c]{@{}c@{}}Systematic Standard \\Uncertainty \boldmath{$b_{\langle X \rangle}$}\end{tabular}}} \\ \midrule
            \multirow{5}{*}{UW} & ATI Mini45-IP65 & $Q$ & $Q_{\mathrm{top}}$ [N-m] & Accuracy: $\pm 1.25\%$ of   full scale load \\ \cmidrule(l){2-5} 
             & ATI Mini40-IP68 & $Q$ & $Q_{\mathrm{bottom}}$ [N-m] & Accuracy: $\pm 1.50\%$ of   full scale load \\ \cmidrule(l){2-5} 
             & Motor Encoder & $\omega$ & $\theta$ [deg.] & Accuracy: $\pm 0.0055$ degrees \\ \cmidrule(l){2-5} 
             & Nortek Vector ADV & $U_{\infty}^3$ & $U_{\infty}$ [m/s] & Accuracy: $\pm 0.5\%$ of   measured value $\pm 1$ mm/s\tnote{\dag} \\ \midrule
            \multirow{4}{*}{UNH} & Interface T8-200 & $Q$ & $Q$ [N-m] & \begin{tabular}[c]{@{}l@{}}Accuracy: $\pm 0.25\%$ of full scale load\\ Nonrepeatability: $\pm 0.05\%$ of full scale load\tnote{$\ddagger$} \end{tabular} \\ \cmidrule(l){2-5} 
             & Kollmoregen AKM62Q & $\omega$ & $\theta$ [deg.] & Accuracy: $\pm 0.0036$ degrees \\ \cmidrule(l){2-5} 
             & Renishaw LM15 & $U_{\infty}^3$ & Tow carriage position [m] & Accuracy: $\pm 100\mu\mathrm{m}$ per meter\tnote{\dag} \\ \bottomrule
            \end{tabular}
            \begin{tablenotes}
                 \item[\dag] $b_{\langle U_{\infty} \rangle}$ converted to $b_{\langle U_{\infty}^3 \rangle}$ by converting the listed value to percentage error of $U_{\infty}$, multiplying by 3 to obtain percentage error of $U_{\infty}^3$, and then multiplying by $U_{\infty}^3$ to convert back to absolute error.
                  \item[$\ddagger$] Accuracy and nonrepeatability combined via root sum of squares.
            \end{tablenotes}
        \end{threeparttable}   
    }
    \label{tab:instrumentSpecs}
\end{table*}

The experimental uncertainty of the measured time-average efficiency, $\langle C_P \rangle$, is evaluated for both the UW and UNH test rigs via the American Society of Mechanical Engineers standard on test uncertainty, ASME PTC 19.1-2005 \citep{american_society_of_mechanical_engineers_test_2005}. Through this procedure, the uncertainty of each component measurement of $\langle C_P \rangle$ (e.g., $\langle Q \rangle$, $\langle \omega \rangle $, $\langle U_{\infty}^3 \rangle$) is estimated, from which the total uncertainty of $\langle C_P \rangle$ is obtained. Uncertainty is evaluated on a cycle-average basis; in other words, the uncertainty of the average value of a given component measurement, $X$, that is obtained for a single rotational cycle is estimated.
The cycle-averages are related to the time-average as
\begin{equation}
    \langle X \rangle = \frac{\sum_{i=1}^{n} X_i}{n} \ \ ,
\end{equation}
\noindent where $n$ is the number of turbine rotational cycles, $X_i$ is the measured average value of $X$ over the $i^{\mathrm{th}}$ cycle, and $\langle X \rangle$ is the average value of $X$ over all cycles, which is equal to the time-average measurement.
Uncertainty is evaluated at the turbine-level rather than at the blade-level since $C_{P,\mathrm{supports}}$ is measured independently from $C_P$. A similar analysis could be conducted to quantify the uncertainty of $\langle C_{P,\mathrm{supports}} \rangle$. However, since common support-structure loss relations were used to compute blade-level efficiency for all tested turbine geometries, this analysis would not inform variability in the trends observed across the parameter space.

\subsection{Method for Estimating Uncertainty}

The following implementation of the standard on test uncertainty is similar to that of \citet{cavagnaro_field_2016} and \citet{snortland_cycletocycle_2023} who have previously applied this method to cross-flow turbine systems.
The uncertainty of each component measurement of $\langle C_P \rangle$ is composed of systematic standard uncertainty and random standard uncertainty. Systematic standard uncertainty, $b_{\langle X \rangle}$, represents the constant error associated with the instruments used to measure $X$. The values of systematic standard uncertainty are estimated for each instrument using the sensor specifications provided by the manufacturers, which are given in \Cref{tab:instrumentSpecs} for each instrument employed in the UW and UNH experimental set-ups.

The random standard uncertainty, $s_{\langle X \rangle}$, represents the variable error associated with repeated measurements of $X$ and is calculated as
\begin{equation}
    s_{\langle X \rangle} = \sqrt{\frac{1}{n (n-1)} \sum_{i=1}^{n} (X_i - \langle X \rangle })^2 \ \ .
\end{equation}
For the UW and UNH experimental set-ups, sources of random standard uncertainty include white electrical noise, Doppler noise, and normally distributed variations in torque, servomotor rotation rate, and inflow or tow carriage velocity.

The combined standard uncertainty of component measurement $X$, $u_{\langle X \rangle}$, is calculated by as
\begin{equation}
    u_{\langle X \rangle} = \sqrt{b_{\langle X \rangle}^2 + s_{\langle X \rangle}^2} \ \ ,
\end{equation}
\noindent and the interval $\langle X \rangle \pm 2u_{\langle X \rangle}$ is expected to contain the true cycle-average value of $X$ with 95\% confidence.

The standard uncertainty for $\langle C_P \rangle$ is obtained by weighting the $b_{\langle X \rangle}$ and $s_{\langle X \rangle}$ of each component measurement by the sensitivity of $\langle C_P \rangle$ to that component. The sensitivity of $\langle C_P \rangle$ to the $j^{\mathrm{th}}$ component measurement is defined as
\begin{equation}
    \label{eq:sensitivity}
    \kappa_{j} = \frac{\partial \langle C_P \rangle}{\partial \langle X_j \rangle} \ \ .
\end{equation}
\noindent The systematic standard uncertainty and random standard uncertainty are then obtained, respectively, as
\begin{equation}
    b_{\langle C_P \rangle } = \sum_{j=1}^{J} \sqrt{(\kappa_{j} b_{\langle X_j \rangle})^2} \ \ ,
\end{equation}
\begin{equation}
    s_{\langle C_P \rangle } = \sum_{j=1}^{J} \sqrt{(\kappa_{j} s_{\langle X_j \rangle})^2} \ \ .
\end{equation}
\noindent where $J$ is the total number of measurement components. Finally, the combined standard error for $\langle C_P \rangle$ is calculated as
\begin{equation}
    u_{\langle C_P \rangle} = \sqrt{b_{\langle C_P \rangle}^2 + s_{\langle C_P \rangle}^2} \ \ ,
\end{equation}
\noindent with the interval $\langle C_P \rangle \pm 2u_{\langle C_P \rangle}$ corresponding to 95\% confidence.

\subsection{Predicted and Measured Uncertainty}

For the experimental test set-up at UW (\Cref{methods:tylerFlume}), the time-average of $C_P$ (\Cref{eq:cp}) can be expressed in terms of its component measurements as
\begin{equation}
    \langle C_P \rangle = \frac{(\langle Q_{\mathrm{top}} \rangle + \langle Q_{\mathrm{bottom}} \rangle) \langle \omega \rangle}{\frac{1}{2} \rho A \langle U_{\infty}^3 \rangle} \ \ ,
\end{equation}
\noindent where $Q_{\mathrm{top}}$ and $Q_{\mathrm{bottom}}$ are the torques measured by the top and bottom load cells, respectively, $\rho$ and $A$ are assumed to be constant, and $\omega(t) \approx \langle \omega \rangle$.
Since under constant speed control the instantaneous angular velocity varies from the time-average angular velocity by at most $\sim\!1\%$ at both facilities, the assumption $\omega(t) \approx \langle \omega \rangle$ is valid and allows the uncertainty contribution of the mechanical power $\langle Q \omega \rangle$ to be decomposed into the individual contributions of $\langle Q_{\mathrm{top}} \rangle$, $\langle Q_{\mathrm{bottom}} \rangle$ and $\langle \omega \rangle$.
However, for other control strategies where both the hydrodynamic torque and angular velocity vary appreciably over the course of a cycle (e.g., intracycle control \citep{strom_intracycle_2017}), the uncertainty of $\langle Q \omega \rangle$ cannot be decomposed in this way.
The sensitivities corresponding to these component measurements are obtained via \Cref{eq:sensitivity} and given as
\begin{equation}
    \label{eq:sens_Q}
    \kappa_{\langle Q_{\mathrm{top}} \rangle} = \kappa_{\langle Q_{\mathrm{bottom}} \rangle } = \frac{\langle C_P \rangle}{\langle Q \rangle} \ \ , 
\end{equation}
\begin{equation}
    \label{eq:sens_omega}
    \kappa_{\langle \omega \rangle} = \frac{\langle C_P \rangle}{\langle \omega \rangle} \ \ , 
\end{equation}
\begin{equation}
    \label{eq:sens_U3}
    \kappa_{\langle U_{\infty}^3 \rangle} = \frac{-\langle C_P \rangle}{\langle U_{\infty}^3 \rangle} \ \ , 
\end{equation}
\noindent where $\langle Q \rangle = \langle Q_{\mathrm{top}} \rangle + \langle Q_{\mathrm{bottom}} \rangle$. 
As described in \Cref{methods:perfMetrics}, at UW measurements of $Q_{\mathrm{top}}$, $Q_{\mathrm{bottom}}$, $\omega$, and turbine position within a rotational cycle are synchronized with each other, whereas measurements of $U_{\infty}$ are acquired separately. Consequently, while average values of $Q$ and $\omega$ can be obtained for each rotational cycle, cycle-average values of $U_{\infty}^3$ cannot be directly calculated. Instead, for $n$ rotational cycles, ``quasi-cycle-average'' values of $U_{\infty}^3$ are calculated by breaking the $U_{\infty}^3$ time-series into $n$ equal-length segments and computing the average of each segment. This provides an estimate of how much average of $U_{\infty}^3$ varied throughout a given test.

For the experimental test set-up at UNH (\Cref{methods:chaseTank}), $\langle C_P \rangle$ can be expressed in terms of its component measurements as
\begin{equation}
    \langle C_P \rangle = \frac{\langle Q \rangle \langle \omega \rangle}{\frac{1}{2} \rho A \langle U_{\infty}^3 \rangle} \ \ ,
\end{equation}
\noindent where once again $\omega(t) \approx \langle \omega \rangle$ under constant speed control. The sensitivities corresponding to $Q$, $\omega$, and $U_{\infty}^3$ are the same as those given in \Cref{eq:sens_Q,eq:sens_omega,eq:sens_U3}. As described in \Cref{methods:perfMetrics}, at UNH measurements of $Q$, $\omega$, $U_{\infty}$, and turbine position within a rotational cycle are synchronized with each other. Therefore, average values of each of these component measurements can be obtained for each rotational cycle.

\begin{figure}[t]
    \centering
    \includegraphics[width=\textwidth]{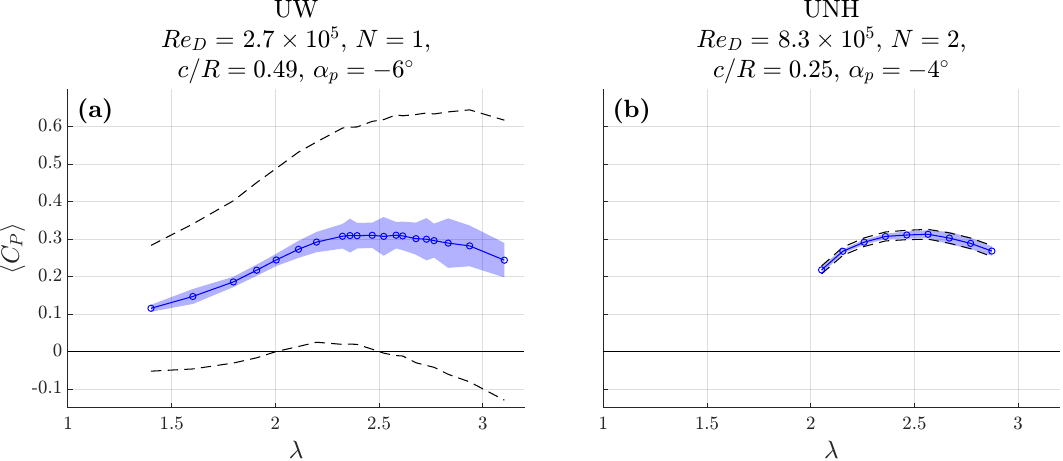}
    {
    \phantomsubcaption\label{fig:cp_uncertainty_uw}
    \phantomsubcaption\label{fig:cp_uncertainty_unh}
    }
    \caption{Uncertainty of $\langle C_P \rangle$ for representative experiments at \subref{fig:cp_uncertainty_uw} UW and \subref{fig:cp_uncertainty_unh} UNH. The dashed lines indicate the 95\% confidence intervals obtained via the standard for test uncertainty, whereas the shaded blue region indicates $\pm 2$ standard deviations of the measured cycle-average values.}
    \label{fig:cp_uncertainty}
\end{figure}

\begin{figure}[t]
    \centering
    \begin{subfigure}[t]{0.45\textwidth}
            \centering
            \includegraphics[width=\textwidth]{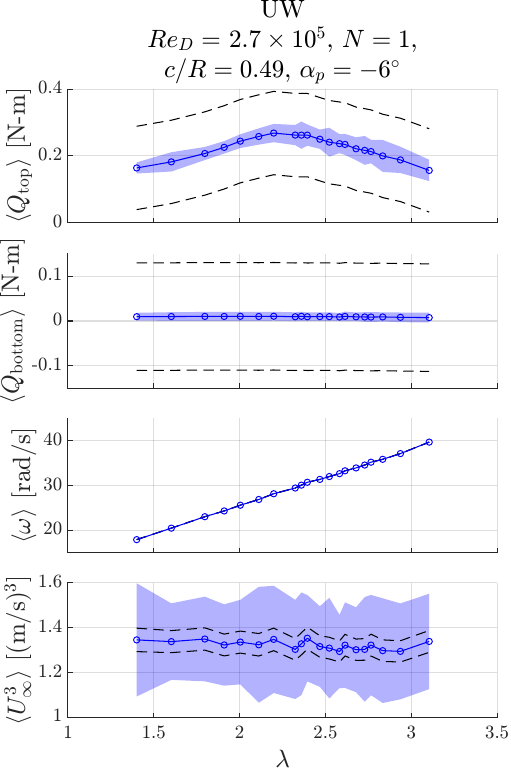}
            \caption{Uncertainties of $\langle Q_\mathrm{top} \rangle$, $\langle Q_\mathrm{bottom} \rangle$, $\langle \omega \rangle$, and $\langle U_{\infty}^3 \rangle$ for a representative experiment at UW (same as in \Cref{fig:cp_uncertainty_uw}).}
            \label{fig:uw_uncertainty}
    \end{subfigure} \hfill
    \begin{subfigure}[t]{0.45\textwidth}
            \centering
            \includegraphics[width=\textwidth]{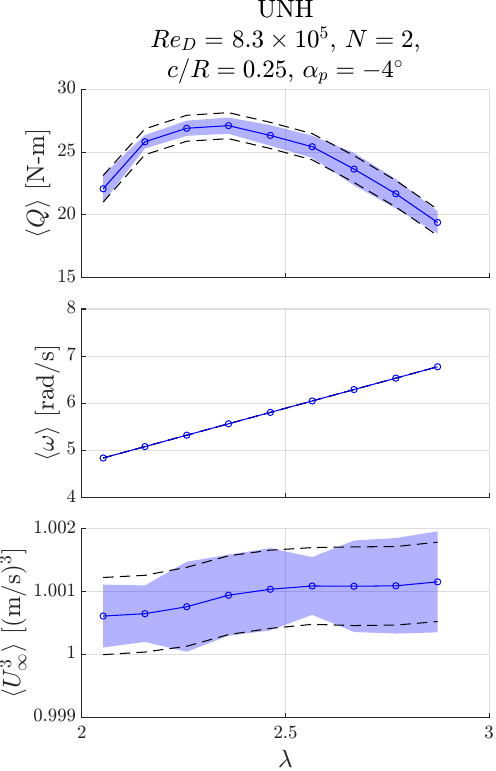}
            \caption{Uncertainties of $\langle Q \rangle$, $\langle \omega \rangle$, and $\langle U_{\infty}^3 \rangle$ for a representative experiment at UNH (same as in \Cref{fig:cp_uncertainty_unh}).}
            \label{fig:unh_uncertainty}
    \end{subfigure}
    \caption{Uncertainties of the component measurements of $\langle C_P \rangle$ for representative experiments at \subref{fig:uw_uncertainty} UW and \subref{fig:unh_uncertainty} UNH. The dashed lines indicate the 95\% confidence intervals obtained via the standard for test uncertainty, whereas the shaded blue region indicates $\pm 2$ standard deviations of the measured cycle-average values. For $\langle \omega \rangle$ at both facilities, both intervals are on the order of the line width shown.}
\end{figure}

\Cref{fig:cp_uncertainty} shows the predicted uncertainty and the observed variability at each point on the $\langle C_P \rangle\!-\!\lambda$ for representative experiments at UW and UNH. 
Across all experiments at UW, the average standard deviation of the cycle-average values of $C_P$ is $\approx\!0.015$ at $\lambda_{\mathrm{opt}}$. As shown in \Cref{fig:cp_uncertainty_uw}, the 95\% confidence interval for $\langle C_P \rangle$ calculated at each $\lambda$ is much larger than the actual range of cycle-average $C_P$ measured at each $\lambda$ during the experiment.

The disparity between predicted uncertainty and observed variability appears to be driven by the relatively large systematic uncertainties for the torques measured by the top and bottom load cells (\Cref{fig:uw_uncertainty}), which show much higher repeatability in practice than what is implied by the manufacturer specifications.
Conversely, larger fluctuations in the quasi-cycle-averaged $U_{\infty}^3$ are observed (\Cref{fig:uw_uncertainty}), which are the result of oscillations in the inflow velocity about the mean as the pumps recirculate water through the flume. These inflow variations likely drive cycle-to-cycle variations in average fluid torque \citep{snortland_cycletocycle_2023} and thus increases cycle-to-cycle variation in $C_P$.
However, the influence of these cycle-to-cycle variations on $\langle C_P \rangle$ is reduced by testing each turbine over many cycles ($\sim \! 50 - 200$ at $\lambda_{\mathrm{opt}}$ across the parameter space) such that convergence to a ``steady-state'' $\langle  C_P \rangle$ is achieved.

Across all experiments at UNH, the average standard deviation of the cycle-average values of $C_P$ is $\approx\!0.005$ at $\lambda_{\mathrm{opt}}$. Compared to the UW results in \Cref{fig:cp_uncertainty_uw}, at UNH there is good agreement between the 95\% confidence interval for $\langle C_P \rangle$ and $\pm2$ standard deviations of the measured cycle-average values (\Cref{fig:cp_uncertainty_unh}). Similar agreement is observed for the component measurements of $\langle C_P \rangle$ (\Cref{fig:unh_uncertainty}).
In general, there is less cycle-to-cycle variability than that observed at UW. This is attributable to the steady freestream velocity provided by the tow carriage controller relative to the UW flume pumps. Consequently, cycle-to-cycle variations in the average fluid torque on the rotor are low, even though the UNH turbines were tested for fewer rotational cycles ($\sim\!10-20$ cycles at $\lambda_{\mathrm{opt}}$) due to the finite tow tank length.

\FloatBarrier

\bibliographystyle{model3-num-names} 
\bibliography{scaling_refs}
\end{document}